\documentclass[10pt,aps,prl,twocolumn,superscriptaddress,longbibliography]{revtex4-1}

\usepackage{epsfig}
\usepackage{dcolumn}
\usepackage{bm}
\usepackage{amsmath}
\usepackage{bbm}
\usepackage{amsfonts}
\usepackage{latexsym}
\usepackage{amssymb}
\usepackage{color}
\usepackage{hyperref}
\usepackage[normalem]{ulem}
\usepackage{soul}

\newcommand{\g}[1]{{\bf #1}}
 
\newcommand{\snte}{Sn$_{1-x}$Pb$_{x}$Te$_{1-y}$Se$_{y}$}

\newcommand{\be}{\begin{equation}}
\newcommand{\ee}{\end{equation}}
\newcommand{\bea}{\begin{eqnarray}}
\newcommand{\eea}{\end{eqnarray}}
\newcommand{\ba}{\begin{eqnarray*}}
\newcommand{\ea}{\end{eqnarray*}}
\newcommand{\dagga}{{\phantom{\dagger}}}

\begin{document}
 
\title{Topological properties of multilayers and surface steps in the SnTe material class} 

\author{Wojciech Brzezicki}
 
\affiliation{International Research Centre MagTop, Institute of Physics, Polish Academy
of Sciences,\\ Aleja Lotnikow 32/46, PL-02668 Warsaw, Poland
} 
 
\author{Marcin M. Wysoki\'nski} 
 
\affiliation{International Research Centre MagTop, Institute of Physics, Polish Academy
of Sciences,\\ Aleja Lotnikow 32/46, PL-02668 Warsaw, Poland
}

\author{Timo Hyart} 
 
\affiliation{International Research Centre MagTop, Institute of Physics, Polish Academy
of Sciences,\\ Aleja Lotnikow 32/46, PL-02668 Warsaw, Poland
}
  

  \begin{abstract}
Surfaces of multilayer semiconductors typically have regions of atomically flat terraces
separated by atom-high steps. Here we investigate the properties of the low-energy states appearing at the
surface atomic steps in \snte. We identify the important approximate symmetries and use them to
construct relevant topological invariants. We calculate the dependence of mirror- and spin-resolved Chern
numbers on the number of layers and show that the step states appear when these invariants are different on
the two sides of the step. Moreover, we find that a particle-hole symmetry can protect one-dimensional Weyl
points at the steps. Since the local density of states is large at the step the system is susceptible to different
types of instabilities, and we consider an easy-axis magnetization as one realistic possibility. We show that
magnetic domain walls support low-energy bound states because the regions with opposite magnetization
are topologically distinct in the presence of non-symmorphic chiral and mirror symmetries, providing a possible
explanation for the zero-bias conductance peak observed in the recent experiment [Mazur {\it et al.}, Phys. Rev. B {\bf 100}, 041408(R) (2019)].
\end{abstract}
\maketitle

\snte\  systems have attracted interest due to realization  of  3D topological crystalline insulator phase   \cite{Fu2011, Hsieh2012, Story2012,Tanaka2012,Hasan2012}, prediction of  2D topological phases  \cite{Liu2015, Safaei2015, Liu2014, Schmidt2018, Schmidt12018} and appearance of  low-energy states at the defects \cite{Fu2014, Sessi2016}. Robust 1D modes were observed at the surface steps separating regions of even and odd number of layers \cite{Sessi2016} and interpreted as topological flat bands using a model obeying a chiral symmetry \cite{Buczko2018}. In a more accurate description  step modes can have a band width but the local density of states (LDOS)  is large so that the system is susceptible to formation of correlated states \cite{Fu2014, Volovik2018, Ojajarvi18}. Recent experiments indicate that an order parameter  emerges at low temperatures and it is accompanied with an appearance of a robust zero-bias peak (ZBCP) in the tunneling conductance \cite{Mazur2017, Das2016}. The temperature and magnetic field dependence of the energy gap are  consistent with superconductivity and under such circumstances the ZBCP is often interpreted as an indication of Majorana  zero modes \cite{Mazur2017, Das2016, Wang2016, Aggarwal2016}, which are intensively searched non-Abelian quasiparticles  \cite{Lutchyn18}. Thus, this  finding calls for a critical study of different mechanisms which may explain the appearance of the ZBCP.

\begin{figure}[h!]
  \begin{center}  
    \includegraphics[width= 0.435\textwidth]{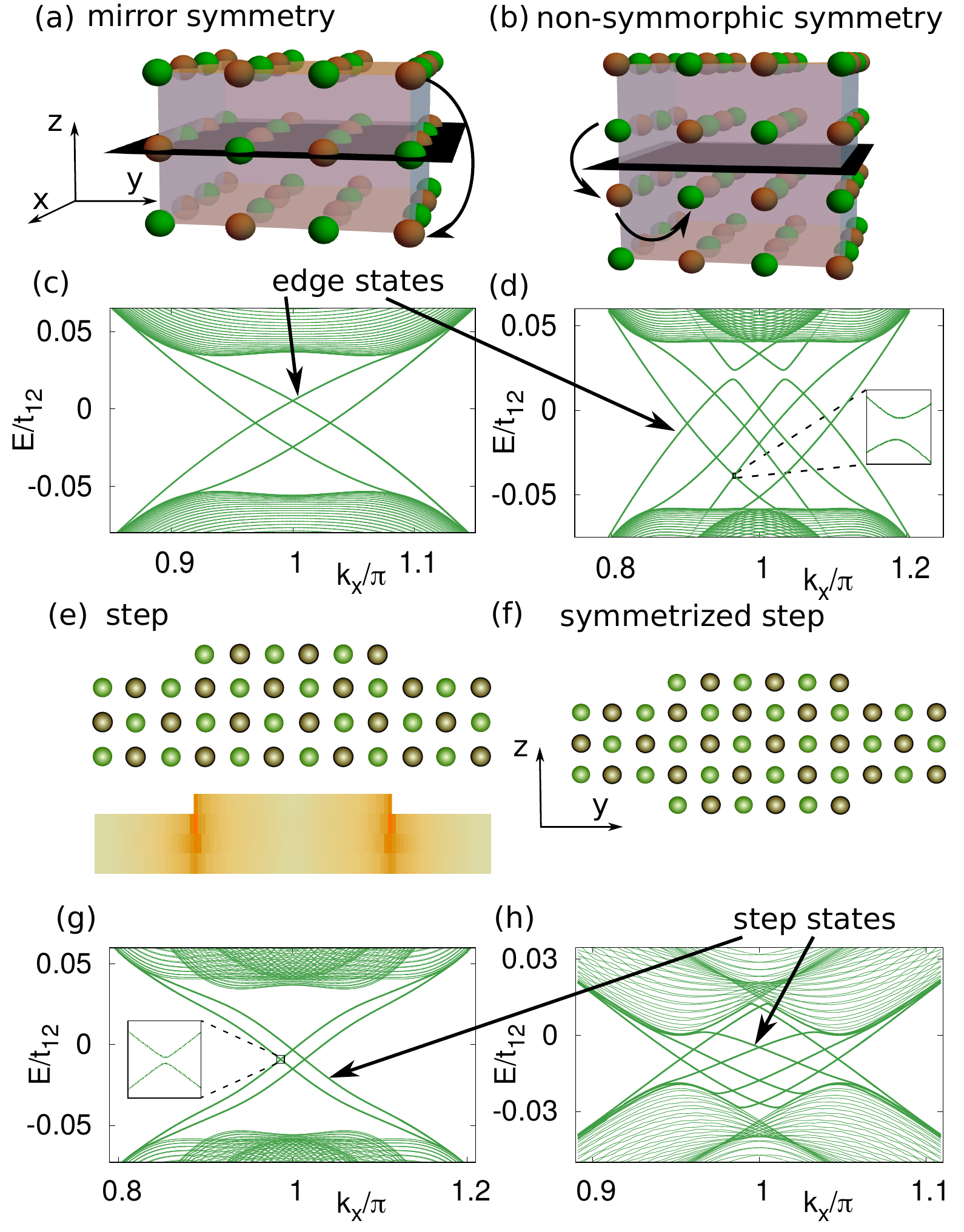}
  \vspace{-0.7	cm}
  \end{center} 
 \caption{(a),(b) Schematic views of the \snte\  multilayers stacked in the (001) direction. Green/brown balls are (Sn,Pb)/(Te,Se) atoms. For odd (even) number of layers $N$ there exists symmorphic (non-symmorphic) mirror symmetry. (c),(d) The corresponding edge-states spectra for $N=3$ and $N=4$.  (e) Surface atomic steps describing an interface between three- and four-layer systems and (f) a symmetrized atomic step. (g),(h) The corresponding spectra for step modes. The panel below (e) shows local density of states of the step states. The width of the sample is $N_y=600$. }
\label{fig1} 
\end{figure}

We show that \snte\ multilayers are a paradigmatic system for realization of  topological phases due to emergent symmetries of the low-energy theory, and ZBCP can appear in the absence of superconductivity. The important 2D topological invariants are the mirror-resolved Chern number $C_{\pm}$ (due to structural mirror symmetry) and spin-resolved Chern number $C_{\uparrow (\downarrow)}$ (due to approximate spin-rotation symmetry). For odd number of layers $N$ the mirror symmetry is a  point-group operation whereas for even $N$ it is a non-symmorphic (NS) symmetry (Fig.~\ref{fig1}), so that adding one layer can change the topology of the system \cite{Schmidt2018,Schmidt12018}. We calculate the dependence of the Chern numbers on $N$ and show that the step states appear when these invariants are different on the two sides of the step. The theory and experiment \cite{Sessi2016,Buczko2018} attribute step states only to odd-height steps, but we predict that also even-height steps can exhibit step-states consistent with other experiment \cite{Iaia2018}.

We discuss the conditions under which a spontaneous symmetry breaking gives rise to an energy gap at the step and study an easy-axis magnetic order as one possibility. We show that magnetic domain walls (DWs) support  low-energy bound states because the regions with opposite magnetization are topologically distinct in the presence of NS chiral and mirror symmetries. Due to the appearance of  DWs the Fermi level is pinned to the energy of the DW states for a range of electron density  providing an explanation for the ZBCP observed in the experiment \cite{Mazur2017}. The observed temperature and magnetic field dependencies   are  consistent with our theory.

Our starting point is a $p$-orbital tight-binding Hamiltonian  describing bulk topological crystalline insulator in \snte-material class \cite{Hsieh2012}
\bea
{\cal H}({\mathbf{k}})&=&m\mathbbm{1}_2\!\otimes\!\mathbbm{1}_3\!\otimes\! \Sigma+t_{12}\!\!\!\sum_{\alpha=x,y,z}\!\!\mathbbm{1}_2
\!\otimes\!\left(\mathbbm{1}_3\!-\!L_{\alpha}^{2}\right)\!\otimes\! h_{\alpha}^{(1)} (k_{\alpha})\nonumber\\
&&\hspace{-0.8cm}+t_{11}\sum_{\alpha\not=\beta}\mathbbm{1}_2\!\otimes\!\left(\mathbbm{1}_3\!-\!\tfrac{1}{2}\left(L_{\alpha}\!+\!\varepsilon_{\alpha\beta}L_{\beta}\right)^{2}\right)
\!\otimes\! h_{\alpha,\beta}^{(2)} (k_{\alpha},k_{\beta})\Sigma \nonumber\\
&&\hspace{-0.8cm} +\sum_{\alpha=x,y,z} \lambda_\alpha \sigma_{\alpha}\!\otimes L_{\alpha}\otimes\mathbbm{1}_8,
\eea
where we have chosen a cubic unit cell with internal sites at the corners labeled by $i=1,\dots,8$ \cite{supplementary}, $\varepsilon_{\alpha\beta}$ is a Levi-Civita symbol, $L_{\alpha}=-i\varepsilon_{\alpha\beta\gamma}$  are the $3\times 3$ angular momentum $L=1$ matrices, $\sigma_\alpha$ are Pauli matrices acting in the spin space, $\Sigma$ is a diagonal $8\times 8$ matrix with entries $s_i=\pm 1$  at the two sublattices [(Sn,Pb)/(Te,Se) atoms],   and $h_{\alpha}^{(1)} (k_{\alpha})$ and $h_{\alpha,\beta}^{(2)} (k_{\alpha},k_{\beta})$  are $8\times 8$ matrices describing hopping between the nearest-neighbor and next-nearest-neighbor lattice sites in the directions $\hat{\alpha}$ and  $\hat{\alpha}+\epsilon_{\alpha \beta} \hat{\beta}$, respectively \cite{supplementary}. We allow the possibility to tune the spin-orbit coupling terms $\lambda_\alpha$ ($\alpha=x,y,z$) to be different from each other although in the real material $\lambda_\alpha=\lambda$. When not otherwise stated
we use $m=1.65$ eV, $t_{12}=0.9$ eV, $t_{11}=0.5$ eV and $\lambda=0.3$ eV \cite{footnote0}.

We focus on Hamiltonian  ${\cal H}_N(\vec{k})$ for $N$ layers in the $z$-direction. The mirror symmetries  for  odd/even N can  be written as $M_z^{o/e}(k_x)=\sigma_{z}\otimes(2L_z^2-1)\otimes m_z^{o/e}(k_x)$, where  $m_z^o$ is a   point-group reflection and $m_z^e(k_x)$  is a (momentum-dependent) NS operation consisting of reflection and a shift by a half lattice vector [Figs.~\ref{fig1}(a,b)]. 
To calculate $C_{\pm}$  we split ${\cal H}_N(\vec{k})$  into two  blocks in the  $M_z^{o/e}(k_x)$ eigenspace. Since $M_z^{o/e}(k_x)$ anticommutes with time-reversal symmetry (TRS) operator ${\cal T}={\cal K}\sigma_y\otimes\mathbbm{1}_3\otimes\mathbbm{1}_8$, these blocks carry opposite Chern numbers $C_{\pm}$ \cite{ChiuRMP}.
We find that $C_+$ oscillates between $+2$ and $-2$ for $N=2n+1$  (we exclude $N=1$) and $C_\pm=0$ for $N=2n$ ($n\in\mathbb{N}$) \cite{supplementary, footnote3}. 
We point out that for even number of layers the non-symmorphic nature of the mirror symmetry guarantees that the Chern number calculated for the blocks must always be equal to zero. Therefore, one can equivalently conclude that the mirror-resolved Chern number does not exist as a topological invariant for even number of layers.
The edge state spectra for $N=3$ and $N=4$ are shown in Figs.~\ref{fig1}(c,d). As predicted by  $C_{\pm}=\mp2$  we see two pairs of  gapless edge modes in the case $N=3$, but surprisingly we find four pairs of edge modes if $N=4$. These edge modes are consistent with $C_{\pm}= 0$ because there exists  small gaps which do not vanish by increasing system size.

\begin{figure}[t]
  \begin{center}  
    \includegraphics[width= 0.45 \textwidth]{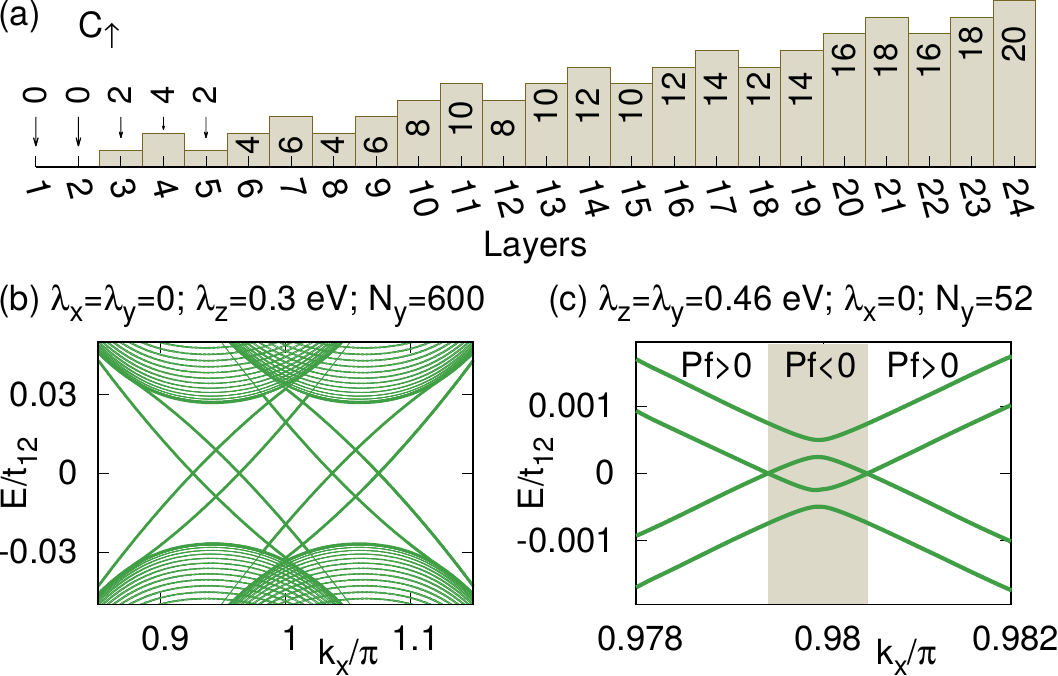}
  \vspace{-0.4	cm}
  \end{center} 
 \caption{(a) Spin-resolved Chern numbers $C_{\uparrow}$ as a function of $N$ for $\lambda_x=\lambda_y=0$. Because we have switched off $\lambda_x$ and $\lambda_y$ the effective spin-orbit coupling becomes smaller. In the case of odd number of layers where $C_\pm$ is a useful topological invariant, we have checked that this reduction of the spin-orbit coupling does not change the $C_\pm$ so that it is reasonable to use $\lambda_z = 0.3 eV$. The only exception is the case $N=3$ where we have used renormalized $\lambda_z=0.5$ eV to keep $C_\pm$ fixed.   (b) The spectrum for $N=4$ showing gapless  edge modes. (c) Gapless spectrum for a step between $N=3$ and $N=4$ can be protected by a  $\mathbb{Z}_2$ invariant if $\lambda_x=0$.}
\label{fig2} 
\end{figure}

To understand the existence of  edge modes in the case of $N=4$ we notice that the momentum in the $z$-direction is quantized and the low-energy degrees of freedom are associated with a motion within the $(x,y)$-plane \cite{footnote1}. Therefore, the components of the spin-orbit coupling $\lambda_\alpha$ ($\alpha=x,y,z$) contribute differently to the spectrum,  and  the dominant effect comes from $\lambda_z\sigma_zL_z$. Hence, turning off $\lambda_x$ and $\lambda_y$  is a good approximation [cf.~Figs.~\ref{fig1}(d) and \ref{fig2}(b)]
and this leads to a spin rotation symmetry with respect to the $z$-axis. Thus, by block-diagonalizing ${\cal H}_N(\vec{k})$ we can calculate $C_{\uparrow}=-C_{\downarrow}$ as a function of $N$ \cite{supplementary}. The results are shown in Fig.~\ref{fig2}(a) and suggest that $C_{\uparrow}$ grows linearly with  $N$ and takes values $C_{\uparrow}=2+4m $ ($m \in \mathbb{Z}$)  for odd $N$ and $C_{\uparrow}=4 m$ ($m \in \mathbb{Z}$) for even $N$.
The careful study whether this behavior persist for arbitrary thickness goes beyond the scope of the current manuscript, but importantly this numerical evidence is already enough that we obtain a topological description of step modes for reasonably large systems, including all the system considered in this manuscript. In particular, the result for $N=4$ is consistent with number of edge modes in Figs.~\ref{fig1}(d), \ref{fig2}(b). For  $\lambda_x=\lambda_y=0$   the tiny gaps originally present in the spectrum vanish completely. 

These Chern numbers provide an interpretation for the appearance of the step modes because they appear whenever  $C_\uparrow$  is different on the two sides of the step [Figs.~\ref{fig1}(e)-(h)] \cite{supplementary}. For a step separating even $N$ and odd $N$ $\Delta C_\uparrow = 2 + 4m$ ($m \in \mathbb{Z}$) and therefore at least two pairs of helical step modes \cite{footnote2}  exist at these steps in agreement with Refs.~\cite{Sessi2016,Buczko2018} [Fig~\ref{fig1}(e),(g)]. 
These step modes are weakly gapped because the spin-rotation symmetry is only present as an approximate symmetry. 
However, we can use $C_+$ to show that these gaps vanish in the limit $N \to \infty$. Namely, by using a symmetrized step-construction shown in Fig.~\ref{fig1}(f),(h) and fixing $N$ so that $C_+=\pm2 $ on the different sides of the step, we find that in a system containing steps on both surfaces there exists $|\Delta C_+|=4$ pairs of gapless edge modes protected by the mirror symmetry. In the limit $N \to \infty$ the step modes at the different surfaces are completely decoupled, and therefore each step supports two pairs of gapless edge modes. 
Interestingly, we find that (depending on $N$) also even-height steps can exhibit $|\Delta C_+| = 4$ or $|\Delta C_\uparrow| = 4|m|$   ($m \in \mathbb{Z}$) pairs of  step modes  consistent with the experiment \cite{Iaia2018}.

In the case of a one atom-high step and finite $N$ the step modes are weakly gapped even though $C_{\pm}$ are different on two sides of the step because the step breaks the mirror symmetry and hybridizes the mirror blocks. However, the step modes can still be exactly gapless for certain widths $N_y$ in $y$-direction if $\lambda_x=0$ [Fig. \ref{fig2}(c)].  This effect comes from the existence of an effective particle-hole symmetry which together with a mirror symmetry gives rise to an antiunitary chiral symmetry  ${\cal S}={\cal K}\sigma_y\otimes (2L_x^2-1)\otimes (i\Sigma m_x)$, where $m_x$ is a mirror reflection with respect to the $x$-plane interchanging the sublattices \cite{supplementary}. Due to this symmetry the Hamiltonian ${\cal H}_{N, N_y}(k_x)$ supports a $\mathbb{Z}_2$ Pfaffian-invariant, which protects a 1D Weyl point  if it changes sign as a function of $k_x$ \cite{Zhao2016, Brzezicki2017}. As demonstrated in Fig.~\ref{fig2}(c) this kind Weyl points can be realized at the steps if $\lambda_x=0$.

\begin{figure*}[t]
  \begin{center}  
    \includegraphics[width= 0.9 \textwidth]{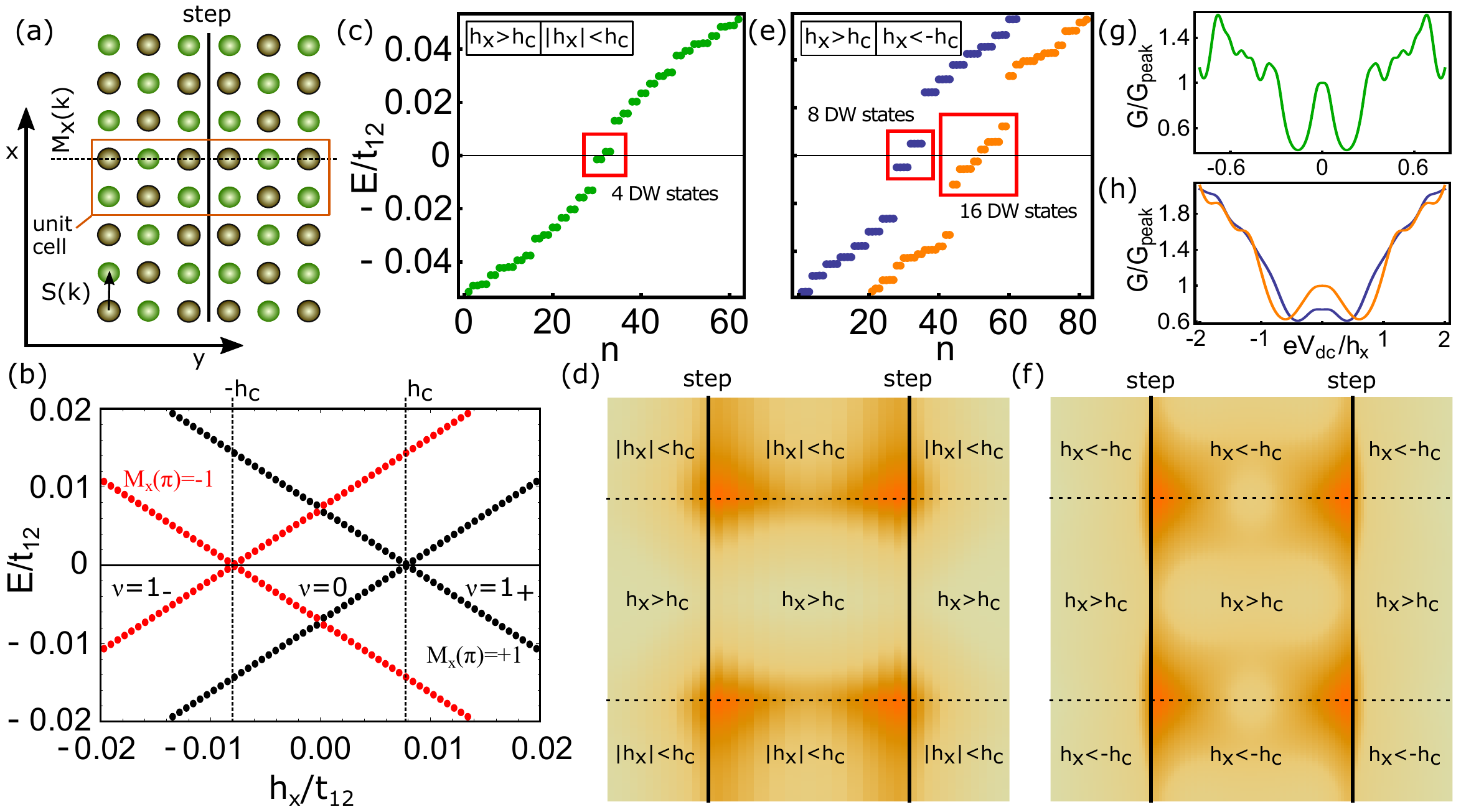}
  \vspace{-0.4	cm}
  \end{center} 
 \caption{(a) Schematic representation of symmetries $M_x(k_x)$ and $S(k_x)$ operating at a single step \cite{supplementary}.  (b) The energies of the step states at $k_x=\pi$ as a function of $h_x$ for $N_y=52$ and $\lambda_z=\lambda_x=0.5$ eV. The states are coloured according to the eigenvalues of $M_x(\pi)$. There exists three topologically distinct phases denoted as trivial phase $\nu=0$ and non-trivial phases with $\nu=1_\pm$. The non-trivial phases are distinct from the trivial phase due to NS chiral $\mathbb{Z}_2$ invariant $\nu$ \cite{supplementary,Shiozaki2015}. The lower index  $\pm 1$ describes the subspace of $M_x(\pi )= \pm 1$ where the band-inversion occurs. (c),(d) Spectrum and LDOS for a system with DWs separating trivial and non-trivial phases. The system supports $4$ low-energy bound states at the 4 DWs. The parameters are $N_x=260$, $N_y=52$ and $\lambda_z=\lambda_x=0.5$ eV. In the trivial (non-trivial) phase $h_x=0.003$ eV ($h_x=0.04$ eV). (e), (f) Same for a system with DWs separating two non-trivial phases with opposite magnetizations. The system supports $8$ low-energy bound states at the 4 DWs [blue dots in (e) and (f)]. If the number of DWs is increased to 8 there exists  $16$ low-energy bound states [orange dots in (e)]. The parameters are $N_x=160$, $N_y=140$, $|h_x|=0.034$ eV,  $\lambda_z=0.5$ eV  and $\lambda_x=0$.  (g),(h) The differential conductance $G$ as a function of bias voltage $V_{dc}$ \cite{supplementary} corresponding to spectra in  (c) and (e), respectively. In all figures $\lambda_y=0$. The magnitude of the gap  and $h_x$  in our simulations are larger than in the experiment by Mazur {\it et al.} \cite{Mazur2017}, but they necessarily decrease for larger $N$ because the bulk gap decreases and the dispersion of the step states gets flatter \cite{supplementary}.}
\label{fig3} 
\end{figure*}

Since we have now established the topological origin of the step modes  in the non-interacting system, we turn our attention to the correlation effects (e.g.~spin, charge, orbital or superconducting order), which are  inevitably present due to the large LDOS in the limit $N \gg 1$ \cite{Fu2014, Volovik2018, Ojajarvi18, supplementary}. Our aim is to show that there exists  a mechanism for the appearance of the ZBCP in the absence of superconductivity (without analyzing the competition between different types of order \cite{supplementary}). We require that the order parameter opens an energy gap, which means that it breaks the symmetries associated with  $C_{\uparrow(\downarrow)}$ and $C_\pm$. Thus,  we assume that there exists a magnetic instability in the vicinity of the steps (due to magnetic impurities or  electron-electron interactions  \cite{supplementary}) giving rise to a Zeeman field $H_Z=\mathbf{h}\cdot \vec{\sigma}$.
Because the step modes are approximately spin-polarized along the $z$-direction the directions of $\mathbf{h}$ within the $(x,y)$-plane are efficient in opening an energy gap, and due to spin-orbit coupling the  gap depends on the direction of $\mathbf{h}$ within the $(x,y)$-plane   \cite{supplementary}. Therefore, the system realizes an easy-axis ferromagnet and the topological defects  are DWs \cite{footnote4}. In the following we consider $\mathbf{h}=(h_x, 0,0)$. 

To study the  topological DW states we determine the low-energy  theory for a single step, the emergent symmetries  and the topological invariants. Although $h_x$ breaks TRS  the spectrum of a system with two steps still exhibits Kramers degeneracy at $k_x=\pi$ [Fig.~\ref{fig1}(g)] due to a remaining NS TRS  ${\cal T}'(k_x)={\cal K}\mathbbm 1_2\otimes (2L_y^2-1)\otimes g(k_x) r_z$,  where  $g(k_x)$ is a diagonal matrix with entries $e^{\pm ikx/2}$  and $r_z$ denotes $\pi$-rotation  with respect to $z$ axis \cite{supplementary}. ${\cal T}'(k_x)$  squares to $+1$ ($-1$) at $k_x=0$  ($k_x=\pi$) which yields Kramer degeneracy only at $k_x=\pi$. We assign half of the states at $k_x=\pi$ to each step by selecting from 
each Kramer's doublet   the state with larger projection on each step. Our low-energy theory is obtained by expanding the Hamiltonian around $k_x=\pi$ in one of the subspaces of the projected states \cite{supplementary}. By assuming $\lambda_y=0$  the system supports a $k$-dependent mirror symmetry $M_x(k_x)=\sigma_x\otimes (2L_x^2-1)\otimes g(k_x)$ and NS chiral symmetry $S(k_x)=i\sigma_y\otimes\mathbbm{1}_3\otimes \Sigma\, m_x  g(k_x)$ [Fig.~\ref{fig3}(a)] \cite{supplementary}. 

We find that in the presence of these symmetries there exists three topologically distinct phases shown in Fig.~\ref{fig3}(b). The trivial phase for $|h_x| < h_c$ is separated from the two non-trivial ones by the energy gap closings at $h_x = \pm h_c$. The non-trivial phases are characterized by a NS chiral $\mathbb{Z}_2$ invariant $\nu =1$  \cite{supplementary, Shiozaki2015}, and  
 the phases at $h_x <-h_c$ and $h_x > h_c$ are topologically distinct because the band-inversions  occur in the different mirror sectors  $M_x(\pi)=\pm 1$ [Fig.~\ref{fig3}(b)]. It is not a priori known whether the DWs between the topologically distinct phases in this symmetry class support DW states \cite{Shiozaki2015}. However, our  calculations show that sharp interfaces  between trivial and non-trivial phases (two non-trivial phases) support one (two) low-energy bound state(s) per DW [Figs.~\ref{fig3}(c-f)]. The number of low-energy states in each case is consistent with the number of zero-energy DW states  expected in the case of smooth DWs with slowly varying spin textures \cite{supplementary}. Although these states resemble the zero-energy DW states considered in the context of Dirac equation \cite{JackiwRebbi} and Su-Schrieffer-Heeger (SSH) model \cite{SSH1, SSH2, SSH3}, there is an important difference because they are realized in a model belonging to a different symmetry class. Namely, the appearance of the DW breaks the symmetries, and therefore the energies of these states in the case of sharp DWs remain non-zero even if the DWs are well-separated  \cite{supplementary}.  Moreover, the energies depend on the tight-binding parameters and in this sense they resemble the topological DW states in systems with more complicated unit cells consisting of three or more atoms \cite{Huda2018arXiv}. Nevertheless, for realistic system parameters the DW states  appear  close to the zero energy \cite{supplementary}.

Because the DW states have non-zero energy, in high-resolution tunneling spectroscopy one would observe two splitted peaks in the  conductance. However, already  small  broadening of the energy levels leads to a single ZBCP [Fig.~\ref{fig3}(g),(h)] \cite{supplementary}. Moreover, for sufficiently large density of DWs the bound states hybridize and form a band inside the energy gap leading to a single ZBCP where the height of the peak depends on the density of DWs [Fig.~\ref{fig3}(e),(h)].  The ZBCP is robust against variations of the density because in analogy to the SSH model \cite{SSH3} we expect that the DWs are the lowest energy charged excitations in the system, and therefore small density of excess electrons (excess holes) is accommodated in the system by increasing the number of DWs, so that up to a critical  variation of the density the Fermi level is pinned to the energy of the DW states. Similar situation occurs in quantum Hall ferromagnets where the lowest energy charged excitations are skyrmions  \cite{Girvin99} which appear due to excess electrons and have been observed experimentally \cite{Barrett95}. Also the parametric dependencies of the ZBCP and the energy gap are  consistent with the experiment \cite{Mazur2017}. The increase of temperature suppresses the order parameter and the energy gap. The external  magnetic field breaks the degeneracy of states with opposite magnetization leading to a confinement between the DWs similarly as a symmetry breaking term in the  SSH model \cite{SSH3}, so that the number of DWs and the magnitude of the ZBCP decrease. Furthermore, we find that by increasing the Zeeman field the energy gap of the system decreases \cite{supplementary}. Therefore, all the observations can be explained without requiring the existence of superconductivity. Finally, the observation that magnetic dopants enhance the ZBCP and the energy gap \cite{Mazur2017} makes it more plausible that the effect originates from magnetic instability instead of superconductivity. The systematic analysis of the correlated states which are consistent with the observations \cite{Mazur2017} and the exploration of the possible common origin of the zero-bias anomalies in various topological semiconductors and semimetals \cite{Mazur2017, Das2016, Wang2016, Aggarwal2016} are interesting directions for future research \cite{supplementary}.
  
\begin{acknowledgments}
We thank T. Dietl, J. Tworzyd\l{}o, \L{}.~Skowronek, G. Mazur, M. Cuoco, R.~Rechci\'nski and R.~Buczko for discussions. 
The work is supported by the Foundation for Polish Science through the IRA
Programme co-financed by EU within SG OP Programme. W.B. also acknowledges support by Narodowe Centrum Nauki (NCN, National
Science Centre, Poland) Project No. 2016/23/B/ST3/00839.
\end{acknowledgments}
%

   %
   

\onecolumngrid

\newpage

\setcounter{equation}{0}

\setcounter{figure}{0}
\setcounter{section}{0}

\section*{Supplementary material for "Topological properties of multilayers and surface steps in the SnTe material class".}   
   
\section{Unit cell, Hamiltonian and symmetries}

\begin{figure}[t]
  \begin{center}  
    \includegraphics[width= 0.5 \textwidth]{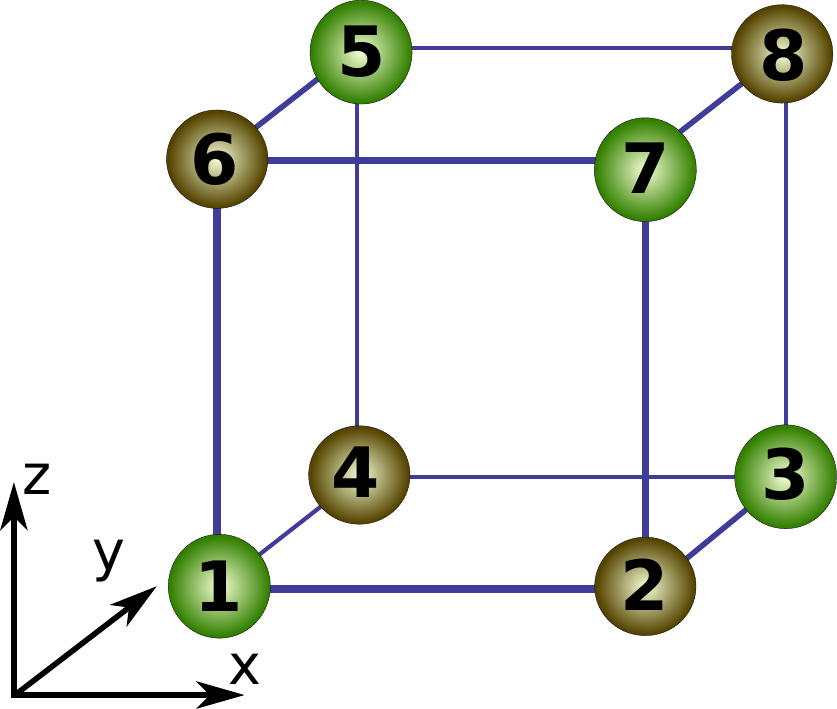}
  \vspace{-0.4	cm}
  \end{center} 
 \caption{The unit cell. The odd end even sites are inequivalent due to opposite mass.}
\label{fig0} 
\end{figure}

Our starting point is a $p$-orbital tight-binding Hamiltonian  describing bulk topological crystalline insulator in \snte-material class \cite{Hsieh2012-supp}
\bea
  H&=&m\sum_j (-1)^j\sum_{\g r, \alpha} \hat c^\dagger_{j\alpha}(\g r)\cdot \hat c^\dagga_{j\alpha}(\g r)+\sum_{j,j'}t_{jj'}\sum_{\langle\g r,\g{r'}\rangle,\alpha} \hat c^\dagger_{j\alpha}(\g r)\cdot \hat d_{\g r\g{r'}}\ \hat d_{\g r\g{r'}}\cdot \hat c^\dagga_{j'\alpha}(\g{r'}) 
  -\sum_j i\lambda \sum_{\g r,\alpha,\beta}\hat c^\dagger_{j\alpha}(\g r)\times \hat c^\dagga_{j\beta}(\g r)\cdot \hat \sigma_{\alpha,\beta}, \nonumber
\eea
where  $\hat c_{j \alpha}(\mathbf{r})$ are vectors of fermionic operators  corresponding to $p_x$-, $p_y$- and $p_z$-orbitals and the indices denote the sublattice $j\in\{1,2\}$  [(Sn,Pb)/(Te,Se) atoms], spin $\alpha$ and lattice site $\mathbf{r}$. Here  $\hat \sigma_{\alpha, \beta}$ is a vector of Pauli matrices, $\hat d_{\g r\g{r'}}$ are unit vectors pointing from  $\g r$ to $\g{r'}$ and the next-nearest-neighbor hoppings satisfy $t_{11}=-t_{22}$.  As discussed in the main text the 3D bulk Hamiltonian can be represented as
\bea
{\cal H}({\mathbf{k}})&=&m\mathbbm{1}_2\!\otimes\!\mathbbm{1}_3\!\otimes\! \Sigma+t_{12}\!\!\!\sum_{\alpha=x,y,z}\!\!\mathbbm{1}_2
\!\otimes\!\left(\mathbbm{1}_3\!-\!L_{\alpha}^{2}\right)\!\otimes\! h_{\alpha}^{(1)} (k_{\alpha})+t_{11}\sum_{\alpha\not=\beta}\mathbbm{1}_2\!\otimes\!\left(\mathbbm{1}_3\!-\!\tfrac{1}{2}\left(L_{\alpha}\!+\!\varepsilon_{\alpha\beta}L_{\beta}\right)^{2}\right)
\!\otimes\! h_{\alpha,\beta}^{(2)} (k_{\alpha},k_{\beta})\Sigma \nonumber\\
&&\hspace{-0.2cm} +\sum_{\alpha=x,y,z} \lambda_\alpha \sigma_{\alpha}\!\otimes L_{\alpha}\otimes\mathbbm{1}_8,
\eea
where $\varepsilon_{\alpha\beta}$ is a Levi-Civita symbol, $L_{\alpha}=-i\varepsilon_{\alpha\beta\gamma}$  are the $3\times 3$ angular momentum $L=1$ matrices and we allow the possibility to tune the spin-orbit coupling terms $\lambda_\alpha$ ($\alpha=x,y,z$) to be different from each other although in the real material $\lambda_\alpha=\lambda$. The matrices $h_{\alpha}^{(1)}(k_{\alpha})$ and $h_{\alpha,\beta}^{(2)}(k_{\alpha}, k_{\beta})$ describe hopping between sites of 8-site unit cells (see Fig.~\ref{fig0}) being 
either nearest or next-nearest neighbors. These are given by,
$$
h_{x}^{(1)}(k_{x})=\begin{pmatrix}0 & 1+e^{-ik_{x}} & 0 & 0 & 0 & 0 & 0 & 0\\
1+e^{ik_{x}} & 0 & 0 & 0 & 0 & 0 & 0 & 0\\
0 & 0 & 0 & 1+e^{ik_{x}} & 0 & 0 & 0 & 0\\
0 & 0 & 1+e^{-ik_{x}} & 0 & 0 & 0 & 0 & 0\\
0 & 0 & 0 & 0 & 0 & 0 & 0 & 1+e^{-ik_{x}}\\
0 & 0 & 0 & 0 & 0 & 0 & 1+e^{-ik_{x}} & 0\\
0 & 0 & 0 & 0 & 0 & 1+e^{ik_{x}} & 0 & 0\\
0 & 0 & 0 & 0 & 1+e^{ik_{x}} & 0 & 0 & 0
\end{pmatrix},
$$
$$
h_{y}^{(1)}(k_{y})=\begin{pmatrix}0 & 0 & 0 & 1+e^{-ik_{y}} & 0 & 0 & 0 & 0\\
0 & 0 & 1+e^{-ik_{y}} & 0 & 0 & 0 & 0 & 0\\
0 & 1+e^{ik_{y}} & 0 & 0 & 0 & 0 & 0 & 0\\
1+e^{ik_{y}} & 0 & 0 & 0 & 0 & 0 & 0 & 0\\
0 & 0 & 0 & 0 & 0 & 1+e^{ik_{y}} & 0 & 0\\
0 & 0 & 0 & 0 & 1+e^{-ik_{y}} & 0 & 0 & 0\\
0 & 0 & 0 & 0 & 0 & 0 & 0 & 1+e^{-ik_{y}}\\
0 & 0 & 0 & 0 & 0 & 0 & 1+e^{ik_{y}} & 0
\end{pmatrix},
$$
$$
h_{z}^{(1)}(k_{z})=\begin{pmatrix}0 & 0 & 0 & 0 & 0 & 1+e^{-ik_{z}} & 0 & 0\\
0 & 0 & 0 & 0 & 0 & 0 & 1+e^{-ik_{z}} & 0\\
0 & 0 & 0 & 0 & 0 & 0 & 0 & 1+e^{-ik_{z}}\\
0 & 0 & 0 & 0 & 1+e^{-ik_{z}} & 0 & 0 & 0\\
0 & 0 & 0 & 1+e^{ik_{z}} & 0 & 0 & 0 & 0\\
1+e^{ik_{z}} & 0 & 0 & 0 & 0 & 0 & 0 & 0\\
0 & 1+e^{ik_{z}} & 0 & 0 & 0 & 0 & 0 & 0\\
0 & 0 & 1+e^{ik_{z}} & 0 & 0 & 0 & 0 & 0
\end{pmatrix},
$$
$$
h_{x,y}^{(2)}=\begin{pmatrix}0 & 0 & 1\!+\!e^{-i(k_{x}+k_{y})} & 0 & 0 & 0 & 0 & 0\\
0 & 0 & 0 & e^{ik_{x}}\!\!+\!e^{-ik_{y}} & 0 & 0 & 0 & 0\\
1+e^{i(k_{x}\!+\!k_{y})} & 0 & 0 & 0 & 0 & 0 & 0 & 0\\
0 & e^{-ik_{x}}\!\!+\!e^{ik_{y}} & 0 & 0 & 0 & 0 & 0 & 0\\
0 & 0 & 0 & 0 & 0 & 0 & e^{-ik_{x}}\!\!+\!e^{ik_{y}} & 0\\
0 & 0 & 0 & 0 & 0 & 0 & 0 & 1\!+\!e^{-i(k_{x}+k_{y})}\\
0 & 0 & 0 & 0 & e^{ik_{x}}\!\!+\!e^{-ik_{y}} & 0 & 0 & 0\\
0 & 0 & 0 & 0 & 0 & 1\!+\!e^{i(k_{x}+k_{y})} & 0 & 0
\end{pmatrix},
$$
$$
h_{y,x}^{(2)}=\begin{pmatrix}0 & 0 & e^{-ik_{x}}\!\!+\!e^{-ik_{y}} & 0 & 0 & 0 & 0 & 0\\
0 & 0 & 0 & 1\!+\!e^{i(k_{x}-k_{y})} & 0 & 0 & 0 & 0\\
e^{ik_{x}}\!\!+\!e^{ik_{y}} & 0 & 0 & 0 & 0 & 0 & 0 & 0\\
0 & 1\!+\!e^{i(k_{y}-k_{x})} & 0 & 0 & 0 & 0 & 0 & 0\\
0 & 0 & 0 & 0 & 0 & 0 & 1\!+\!e^{i(k_{y}-k_{x})} & 0\\
0 & 0 & 0 & 0 & 0 & 0 & 0 & e^{-ik_{x}}\!\!+\!e^{-ik_{y}}\\
0 & 0 & 0 & 0 & 1\!+\!e^{i(k_{x}-k_{y})} & 0 & 0 & 0\\
0 & 0 & 0 & 0 & 0 & e^{ik_{x}}\!\!+\!e^{ik_{y}} & 0 & 0
\end{pmatrix},
$$
$$
h_{y,z}^{(2)}=\begin{pmatrix}0 & 0 & 0 & 0 & 1\!+\!e^{-i(k_{y}+k_{z})} & 0 & 0 & 0\\
0 & 0 & 0 & 0 & 0 & 0 & 0 & 1\!+\!e^{-i(k_{y}+k_{z})}\\
0 & 0 & 0 & 0 & 0 & 0 & e^{ik_{y}}\!\!+\!e^{-ik_{z}} & 0\\
0 & 0 & 0 & 0 & 0 & e^{ik_{y}}\!\!+\!e^{-ik_{z}} & 0 & 0\\
1\!+\!e^{i(k_{y}+k_{z})} & 0 & 0 & 0 & 0 & 0 & 0 & 0\\
0 & 0 & 0 & e^{-ik_{y}}\!\!+\!e^{ik_{z}} & 0 & 0 & 0 & 0\\
0 & 0 & e^{-ik_{y}}\!\!+\!e^{ik_{z}} & 0 & 0 & 0 & 0 & 0\\
0 & 1\!+\!e^{i(k_{y}+k_{z})} & 0 & 0 & 0 & 0 & 0 & 0
\end{pmatrix},
$$
$$
h_{z,y}^{(2)}=\begin{pmatrix}0 & 0 & 0 & 0 & e^{-ik_{y}}\!\!+\!e^{-ik_{z}} & 0 & 0 & 0\\
0 & 0 & 0 & 0 & 0 & 0 & 0 & e^{-ik_{y}}\!\!+\!e^{-ik_{z}}\\
0 & 0 & 0 & 0 & 0 & 0 & 1\!+\!e^{i(k_{y}-k_{z})} & 0\\
0 & 0 & 0 & 0 & 0 & 1\!+\!e^{i(k_{y}-k_{z})} & 0 & 0\\
e^{ik_{y}}\!\!+\!e^{ik_{z}} & 0 & 0 & 0 & 0 & 0 & 0 & 0\\
0 & 0 & 0 & 1\!+\!e^{i(k_{z}-k_{y})} & 0 & 0 & 0 & 0\\
0 & 0 & 1\!+\!e^{i(k_{z}-k_{y})} & 0 & 0 & 0 & 0 & 0\\
0 & e^{ik_{y}}\!\!+\!e^{ik_{z}} & 0 & 0 & 0 & 0 & 0 & 0
\end{pmatrix},
$$
$$
h_{x,z}^{(2)}=\begin{pmatrix}0 & 0 & 0 & 0 & 0 & 0 & 1\!+\!e^{-i(k_{x}+k_{z})} & 0\\
0 & 0 & 0 & 0 & 0 & e^{ik_{x}}\!\!+\!e^{-ik_{z}} & 0 & 0\\
0 & 0 & 0 & 0 & e^{ik_{x}}\!\!+\!e^{-ik_{z}} & 0 & 0 & 0\\
0 & 0 & 0 & 0 & 0 & 0 & 0 & 1\!+\!e^{-i(k_{x}+k_{z})}\\
0 & 0 & e^{-ik_{x}}\!\!+\!e^{ik_{z}} & 0 & 0 & 0 & 0 & 0\\
0 & e^{-ik_{x}}\!\!+\!e^{ik_{z}} & 0 & 0 & 0 & 0 & 0 & 0\\
1\!+\!e^{i(k_{x}+k_{z})} & 0 & 0 & 0 & 0 & 0 & 0 & 0\\
0 & 0 & 0 & 1\!+\!e^{i(k_{x}+k_{z})} & 0 & 0 & 0 & 0
\end{pmatrix},
$$
$$
h_{z,x}^{(2)}=\begin{pmatrix}0 & 0 & 0 & 0 & 0 & 0 & e^{-ik_{x}}\!\!+\!e^{-ik_{z}} & 0\\
0 & 0 & 0 & 0 & 0 & 1\!+\!e^{i(k_{x}-k_{z})} & 0 & 0\\
0 & 0 & 0 & 0 & 1\!+\!e^{i(k_{x}-k_{z})} & 0 & 0 & 0\\
0 & 0 & 0 & 0 & 0 & 0 & 0 & e^{-ik_{x}}\!\!+\!e^{-ik_{z}}\\
0 & 0 & 1\!+\!e^{i(k_{z}-k_{x})} & 0 & 0 & 0 & 0 & 0\\
0 & 1\!+\!e^{i(k_{z}-k_{x})} & 0 & 0 & 0 & 0 & 0 & 0\\
e^{ik_{x}}\!\!+\!e^{ik_{z}} & 0 & 0 & 0 & 0 & 0 & 0 & 0\\
0 & 0 & 0 & e^{ik_{x}}\!\!+\!e^{ik_{z}} & 0 & 0 & 0 & 0
\end{pmatrix}.
$$
$\Sigma$ is a diagonal matrix describing mass modulation in a unit cell and has
diagonal entries $(-1,1,-1,1,-1,1,-1,1 )$. 
For $\lambda_\alpha=0$  the system, which is translationally invariant in at
least $x$ direction, has a NS chiral symmetry of the form $S_{\alpha}(k_x)=i\sigma_{\alpha}\otimes\mathbbm{1}_3\otimes \Sigma\, m_x  g(k_x)$,
where $m_x$ is a mirror-$x$ reflection of the unit cell given by
$$
m_{x}=\begin{pmatrix}0 & 1 & 0 & 0 & 0 & 0 & 0 & 0\\
1 & 0 & 0 & 0 & 0 & 0 & 0 & 0\\
0 & 0 & 0 & 1 & 0 & 0 & 0 & 0\\
0 & 0 & 1 & 0 & 0 & 0 & 0 & 0\\
0 & 0 & 0 & 0 & 0 & 0 & 0 & 1\\
0 & 0 & 0 & 0 & 0 & 0 & 1 & 0\\
0 & 0 & 0 & 0 & 0 & 1 & 0 & 0\\
0 & 0 & 0 & 0 & 1 & 0 & 0 & 0
\end{pmatrix},
$$
and $g(k_x)$ is a diagonal gauge matrix with the entries $(e^{i\frac{k_{x}}{2}},e^{-i\frac{k_{x}}{2}}, e^{-i\frac{k_{x}}{2}},e^{i\frac{k_{x}}{2}},e^{i\frac{k_{x}}{2}},e^{i\frac{k_{x}}{2}},e^{-i\frac{k_{x}}{2}},e^{-i\frac{k_{x}}{2}} )$. The plane of the mirror $m_x$ cuts the unit cell into halves and interchanges the sublattices and this is why it appears in the chiral symmetry operator. Its relation to the Hamiltonian is $\{{\cal H}({\mathbf{k}}),S_{\alpha}(k_x)\}=0$ for every $k$-point. In the main text we always use $S(k_x)\equiv S_{y}(k_x)$ so the $y$ subscript is omitted for brevity. We also point out that the sign of $\lambda$ is not important for our results because we can construct a chiral symmetry $S_0(k_x)$ by using the identity matrix $\sigma_0$ in the expression given above, such that the Hamiltonians with positive and negative spin-orbit coupling are related to each other as ${\cal H}_\lambda(\mathbf{k}) S_0(k_x)=-S_0(k_x) {\cal H}_{-\lambda}(\mathbf{k})$.
 The same gauge $g(k_x)$ matrix is used for constructing the  $x$-plane mirror reflection in the $k$-space $M_x(k_x)=\sigma_x\otimes (2L_x^2-1)\otimes g(k_x)$. It satisfies the usual relation with the Hamiltonian $M_x(k_x){\cal H}(k_x,k_y,k_z)M_x^{\dagger}(k_x)={\cal H}(-k_x,k_y,k_z)$.
Analogically the  $z$-plane mirror reflection $M_z(k_x)$ can be defined as $M_z(k_x)=\sigma_z\otimes (2L_z^2-1)\otimes g(-k_x)m_z$, where $m_z$ transforms the sites in the unit cell with a matrix
$$
m_{z}=\begin{pmatrix}0 & 0 & 0 & 0 & 0 & 0 & 1 & 0\\
0 & 0 & 0 & 0 & 0 & 1 & 0 & 0\\
0 & 0 & 0 & 0 & 1 & 0 & 0 & 0\\
0 & 0 & 0 & 0 & 0 & 0 & 0 & 1\\
0 & 0 & 1 & 0 & 0 & 0 & 0 & 0\\
0 & 1 & 0 & 0 & 0 & 0 & 0 & 0\\
1 & 0 & 0 & 0 & 0 & 0 & 0 & 0\\
0 & 0 & 0 & 1 & 0 & 0 & 0 & 0
\end{pmatrix}
$$
and we get $M_z(k_x){\cal H}(k_x,k_y,k_z)M_z^{\dagger}(k_x)={\cal H}(k_x,k_y,-k_z)$.
Finally, there exists also a symmorphic $\pi$-rotation symmetry $R_z$ in the $xy$ plane $R_z{\cal H}(k_x,k_y,k_z)R_z^{\dagger}={\cal H}(-k_x,-k_y,k_z)$, where $R_z=\sigma_z\otimes (2L_z^2-1)\otimes r_z$ and
$$
r_{z}=\begin{pmatrix}0 & 0 & 1 & 0 & 0 & 0 & 0 & 0\\
0 & 0 & 0 & 1 & 0 & 0 & 0 & 0\\
1 & 0 & 0 & 0 & 0 & 0 & 0 & 0\\
0 & 1 & 0 & 0 & 0 & 0 & 0 & 0\\
0 & 0 & 0 & 0 & 0 & 0 & 1 & 0\\
0 & 0 & 0 & 0 & 0 & 0 & 0 & 1\\
0 & 0 & 0 & 0 & 1 & 0 & 0 & 0\\
0 & 0 & 0 & 0 & 0 & 1 & 0 & 0
\end{pmatrix}.
$$
By combining $R_z$ with $M_x(k_x)$ we can obtain a $y$-plane mirror reflection $M_y(k_x)=R_zM_x(k_x)$ satisfying  $M_y(k_x){\cal H}(k_x,k_y,k_z)M_y(k_x)^{\dagger}={\cal H}(k_x,-k_y,k_z)$,
and by combining $R_z$ with $M_z(k_x)$ we get an inversion symmetry $I(k_x)=R_zM_z(k_x)$ satisfying $I(k_x){\cal H}({\mathbf{k}})I^{\dagger}(k_x)={\cal H}(-{\mathbf{k}})$. The inversion symmetry can be  combined with time-reversal symmetry ${\cal T}={\cal K}\sigma_y\otimes\mathbbm{1}_3\otimes\mathbbm{1}_8$ where ${\cal K}$ is complex conjugation. This way we obtain an operator ${\cal Q}(k_x)={\cal T}I(k_x)$ satisfying ${\cal Q}(k_x){\cal H}({\mathbf{k}}){\cal Q}^{-1}(k_x)={\cal H}({\mathbf{k}})$. Since ${\cal Q}^2(k_x)=-1$ for all $k_x$, every band is Kramers degenerate within the whole Brillouin zone (BZ).  

When the system is not fully 3D but consists of a finite number of atomic layers $N$ stacked in the $z$ direction the spatial part of the symmetry operators may become richer if they permute the layers. When $N$ is even the supercell can be constructed by stacking cubic cells vertically  but in case of odd $N$ we have to stack $(N-1)/2$ of whole cubic cells and one half extra.  This means that for even $N$ the $z$-plane mirror reflection is a NS symmetry $M_z^e(k_x)$, where the intracell reflection $m_z$ is otherwise the same as in the case of bulk $M_z(k_x)$ but it also reverts the order of the unit cells in $z$ direction. On the other hand,  for odd $N$ the reflection plane coincides with the central layer of the system, and this leads to a symmorphic $M_z^o$ version of the $M_z(k_x)$ operator (with no $k_x$ dependence). 

In both even and odd case  $M_z^{o,e}(k_x)$ anticommutes with ${\cal T}$, i.e.~${\cal T} M_z^{o,e}(k_x)+M_z^{o,e}(-k_x){\cal T}=0$, and commutes with the multilayer 2D Hamiltonian $[{\cal H}_N(k_x,k_y),M_z^{o,e}(k_x)]=0$. This implies that the double degeneracy of every band is resolved in the $M_z^{o,e}(k_x)$ invariant subspaces. It is due to the peculiarity of the NS symmetry that in the case of even $N$  the  $M_z^e(k_x)$ subspaces keep internal TRS. To understand this \cite{Brzezicki2017s} we can first order the eigenvectors of $M_z^e(k_x)$ based on the eigenvalues and write them in the columns of matrix $U(k_x)$. Then, in the basis given by $U(k_x)$ the mirror symmetry $M_z^e(k_x)$ is diagonal, and we denote the Hamiltonian and TRS  as

\be\widetilde{{\cal H}}_N(k_x,k_y)=U(k_x)^{\dagger}{\cal H}_N(k_x,k_y)U(k_x), \quad \widetilde{{\cal T}}={\cal K}U(k_x)^{T}(\sigma_y\otimes\mathbbm{1}_3\otimes\mathbbm{1}_8)U(k_x).\ee

The Hamiltonian commutes with  $M_z^e(k_x)$ so it has block-diagonal form but TRS is block-off-diagonal due to anticommutation with $M_z^e(k_x)$. Apparently there is no TRS inside the diagonal blocks of 
$\tilde{{\cal H}}_N$. One can however notice that $U(k_x)$ is not $2\pi$- but $4\pi$-periodic due to nonsymmorphicity so one can define a non-trivial unitary matrix $\chi$ that produces a $2\pi$ shift of the momentum in the $U(k_x)$ basis

\be\chi=U(k_x+2\pi)^{\dagger}U(k_x), \quad\quad \chi\widetilde{{\cal H}}_N(k_x,k_y)\chi^{\dagger}=\widetilde{{\cal H}}_N(k_x+2\pi,k_y).\ee

Now, knowing that eigensubspaces of  $M_z^e(k_x)$ are interchanged after a shift of $2\pi$ it is easy to notice that $\chi$ has a purely block-off-diagonal structure, just like $\widetilde{{\cal T}}$. Therefore,  by combining these two operators we obtain an intra-block TRS, i.e.,

\be\widetilde{{\cal T}}_{2}=\widetilde{{\cal T}}\chi, \quad\quad\widetilde{{\cal T}}_{2}\widetilde{{\cal H}}_N(k_x,k_y)\widetilde{{\cal T}}_{2}^{-1}=\widetilde{{\cal H}}_N(-k_x-2\pi,-k_y),\ee
which is a time-reversal with respect to the $(k_x,k_y)=(-\pi,0)$ point. This relation is enough to guarantee that the mirror-resolved Chern numbers vanish in the case of even number of layers.
Another way of resolving the global double degeneracy is to have a spin-rotation symmetry around some axis which happens when only one component of SOC is present, i.e., having only $\lambda_z$ non-zero yields a symmetry given by $S_z=\sigma_{z}\otimes\mathbbm{1}_3\otimes\mathbbm{1}_8$. The spin subspaces do not depend on spatial symmetries so the spin-resolved Chern numbers are allowed for any $N$.

\begin{figure}[h!]
  \begin{center}  
    \includegraphics[width=  \textwidth]{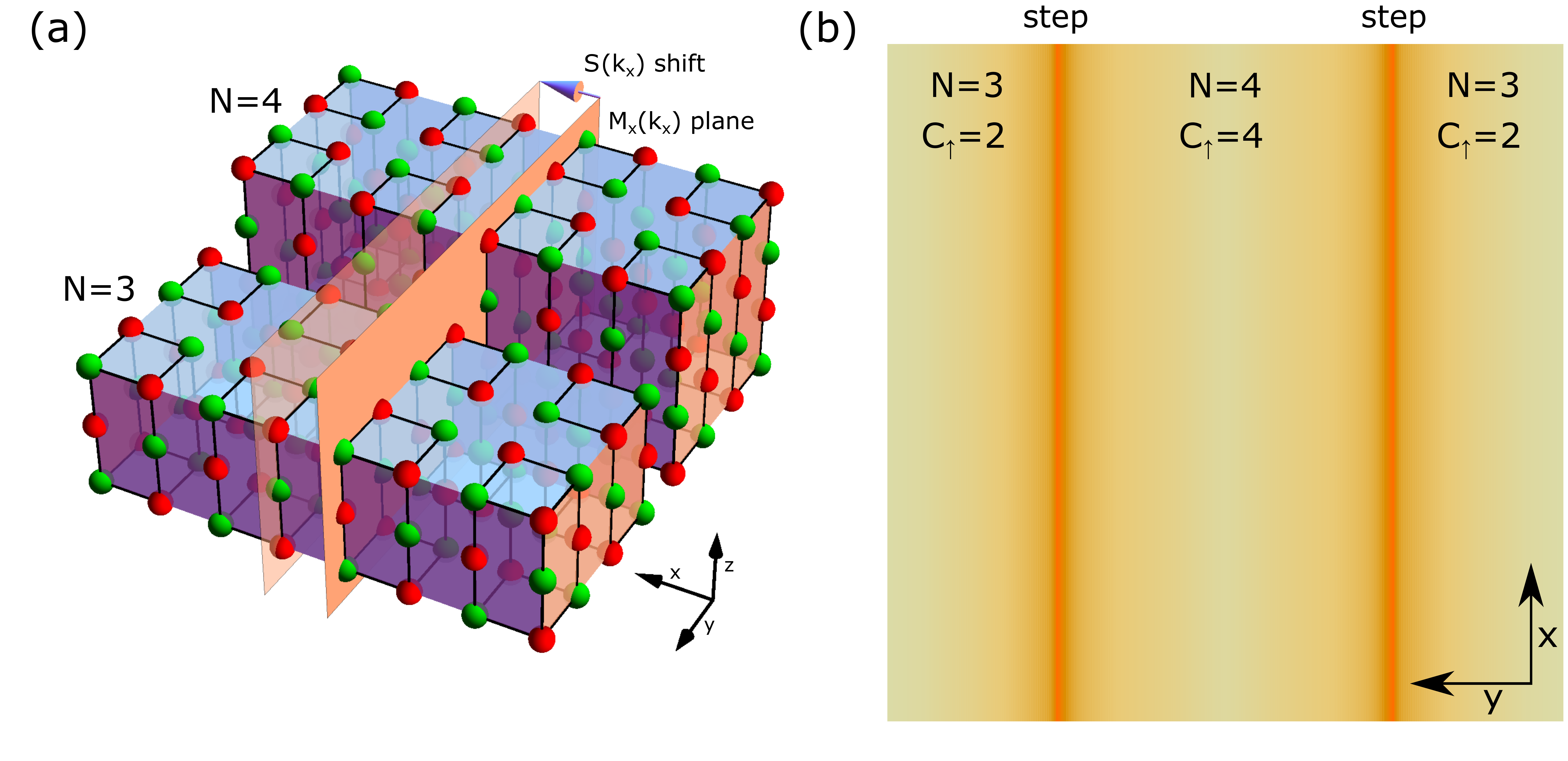}
  \vspace{-1.3	cm}
  \end{center} 
 \caption{(a) Schematic view of a surface atomic step formed by interfacing systems with $N=3$ and $N=4$ layers. Unit cells for both systems are marked with black lines. The mirror plane related to the $M_x(k_x)$ symmetry and the shift vector related to the  chiral symmetry $S(k_x)$ are shown. (b) Local density of states corresponding to the low-energy step states for a similar system containing two surface atomic steps. The width of the system is $N_y=600$ and we have used parameters $m=1.65$ eV, $t_{12}=0.9$ eV, $t_{11}=0.5$ eV and $\lambda=0.3$ eV. Spin-resolved Chern numbers $C_{\uparrow}$ of both subsystems are marked in the figure.}
\label{sfig9} 
\end{figure}

When a step along $x$ is introduced in the system, see Fig. \ref{sfig9}(a),  the dimension of the $k$-space is further reduced to 1D and we denote the Hamiltonian as ${\cal H}_{\{N,N'\}}(k_x)$ where $N$ and $N'$ denote the number of layers at the either side of the step edge. Note that apart from the effective Hamiltonian obtained by projecting the Hamiltonian to a single step we always considered systems which are periodic in the $y$ direction, so that the steps come in pairs and the $y$-plane mirror $M_y(k_x)$ is conserved.
Then for $\lambda_x=0$ we can define an antiunitary  chiral symmetry ${\cal S}={\cal T}M_x(k_x)S_x(k_x)$ which is symmorphic and satisfies ${\cal S}^2=+1$. It allows for transformation that makes the Hamiltonian purely imaginary. Its explicit form is ${\cal S}={\cal K}\sigma_y\otimes (2L_x^2-1)\otimes (i\Sigma m_x)$ and the relation with the Hamiltonian is usual anticommutation although due to antiunitarity it does not yield two off-diagonal blocks of ${\cal H}_{\{N,N'\}}(k_x)$. A part of ${\cal S}$ symmetry is ${\cal T}S_x(k_x)$ which is referred as an effective particle-hole symmetry in the main text. What is peculiar about ${\cal S}$ is that it is compatible with a Zeeman field term in any direction.
Finally, if we assume  $\mathbf{h}=(h_x, 0,0)$ and $\lambda_y=0$ the unitary symmetries that survive are mirror $M_x(k_x)$ and chirality $S_y(k_x)\equiv S(k_x)$. 
The $y$-plane mirror $M_y(k_x)$ is not a symmetry anymore because it anticommutes with the Zeeman field term but together with  ${\cal T}$ it forms a NS time-reversal symmetry ${\cal T}'(k_x)={\cal T}M_y(k_x)$ that can be expressed as ${\cal T}'(k_x)={\cal K}\mathbbm 1_2\otimes (2L_y^2-1)\otimes g(k_x)  r_z$ and it satisfies ${\cal T}'(k_x){\cal H}_{\{N,N'\}}(k_x){\cal T}'^{-1}(k_x)={\cal H}_{\{N,N'\}}(-k_x)$. It is another peculiarity of a NS symmetry that ${\cal T}'^2(0)=1$ but ${\cal T}'^2(\pi)=-1$ so that the Kramers degeneracy appears only at $k_x=\pi$ \cite{Brzezicki2017s}. 
 
\section{Calculation of the mirror- and spin-resolved Chern numbers}

Chern number for a given band can be calculated from the definition involving Berry connection and Berry curvature.
This can  be quite non-trivial when the bands are degenerate \cite{Gradhand_2012}
but for the simple case of non-degenerate bands one defines the Berry connection for the $n$-th band as
\be
\vec{A}_{n}=i\left\langle n,\vec{k}\right|\vec{\nabla}_{\vec{k}}\left|n,\vec{k}\right\rangle, 
\ee
and the Berry curvature as 
\be
\Omega_{n}=\vec{\nabla}_{\vec{k}}\times\vec{A_{n}}.
\ee
The integral of the Berry curvature over the Brillouin zone summed over all occupied bands  gives an integer called Chern number 
\be
{\cal C}=\frac{1}{2\pi}\underset{BZ}{\int}d^{2}k\sum_{n\le n_{F}}\Omega_{n}.
\ee
This definition of Chern number is however not convenient for numerical calculation because it requires a smooth gauge of 
vectors $\left|n,\vec{k}\right\rangle$ which for large matrices are known only numerically. This difficulty can be overcome by
special gauge-invariant prescriptions for Berry curvature \cite{Fukui2005,Deng2012} but since they yield Berry curvature band by
band they become singular whenever there is a band crossing.  If we do not need to know the curvatures of individual 
bands, and this is the case here, the most practical way is to express the Chern number as
\be
{\cal C}=\frac{1}{\pi}\underset{BZ}{\int}d^{2}k\sum_{{n\le n_{F}\atop n'>n_{F}}}\mathfrak{Im}\frac{\left\langle n,\vec{k}\right|\partial_{k_{x}}{\cal H}\left|n',\vec{k}\right\rangle \left\langle n',\vec{k}\right|\partial_{k_{y}}{\cal H}\left|n,\vec{k}\right\rangle }{\left(E_{\vec{k}}^{(n)}-E_{\vec{k}}^{(n')}\right)^{2}},
\ee
where in the sum $n$ runs over the occupied bands and  $n'$ over unoccupied ones.  The above equation is so-called  Kubo formula 
that describes Hall conductivity in the quantum Hall systems \cite{Nagaosa2010}.
In our case we do not insert the full Hamiltonian into the Kubo formula because it would give zero Chern number due to TRS. Instead we insert one of the  
diagonal blocks of ${\cal H}$ in the eigenbasis of $M^{o,e}_z(k_x)$  for the mirror-resolved Chern number and $S_z$ for the spin-resolved Chern number.
The resulting mirror-resolved Chern numbers $C_+$ are given in the Table \ref{tab1}. We note that they keep on oscillating between values $\pm 2$
for $N>1$ as the number of layers grows. Assuming that this behavior persists for arbitrary $N$ we predict that the step states appear generically in the limit
of large $N$ when every surface step can be well approximated by a  symmetrized step [Fig. 1(f) of the main text]. 

\begin{table}[h]
\caption{\label{tab1}
Mirror Chern numbers $C_+$ for systems with odd number of layers $N$ ($C_+=0$ for even $N$). Parameters are $m=1.65$ eV, $t_{12}=0.9$ eV, $t_{11}=0.5$ eV , $\lambda_x=\lambda_y=\lambda_z=0.3$ eV. 
}
\begin{tabular}{|@{\quad}  l @{\quad}|@{\quad} c @{\quad}|}
\colrule
\textrm{$N$}&
\textrm{$C_+$}\\
\colrule
1 & 0\\
3 &-2\\
5 & 2\\
7 & 2\\
9 & -2\\
11 &-2\\
13 & 2\\
15 & 2\\
17 & -2\\
19 & -2\\
\colrule
\end{tabular}
\end{table}


\section{Approximate spin-rotation symmetry versus topology}

\begin{figure}[b!]
  \begin{center}  
    \includegraphics[width=  0.5 \textwidth]{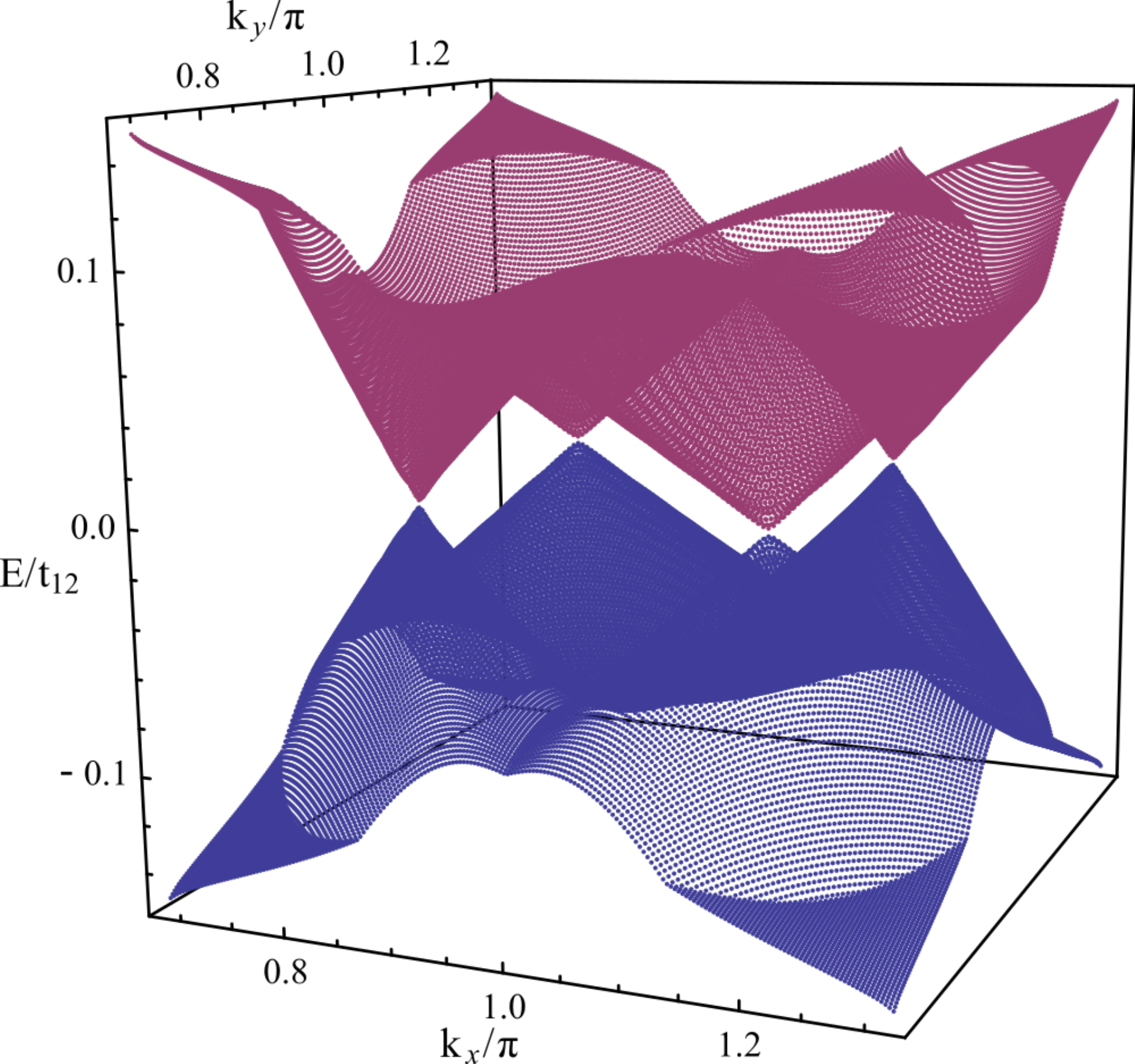}
  \vspace{0.0	cm}
  \end{center} 
 \caption{Surface Dirac cones for $N=200$ and parameters set as  $m=1.65$ eV, $t_{12}=0.9$ eV, $t_{11}=0.5$ eV , $\lambda_z=0.3$ eV and $\lambda_x=\lambda_y=0$. }
\label{sfig10} 
\end{figure}

In the main text we argue that taking $\lambda_x=\lambda_y=0$ is a good approximation for the low-energy theory of a system where the layers are stacked along the $z$-direction. Namely, because of the  confinement in the $z$ direction  the $p_x$- and $p_y$-orbitals are active and thus $\lambda_z$  is responsible for the dominating spin-orbit coupling term.  One can however ask whether this is still a good approximation for a very thick multilayer that seems to approach the three-dimensional case. In such a case the non-trivial topology is guaranteed by the mirror Chern number in the diagonal mirror planes, for instance in the $(110)$ plane. If all components of the spin-orbit coupling are present this invariant for the parameters used in the main text takes values ${\cal C}_{\pm}=\pm 2$ and leads to four Dirac points around X point at the $(001)$ surface of the system.  Now we can ask what happens if we assume conserved  spin-rotation symmetry. Is it possible that this will not affect the topology of a 3D system? The answer is positive if we choose the axis of the spin-rotation symmetry to be perpendicular to the $(110)$ plane when we calculate the mirror Chern number i.e.~we obtain the same mirror Chern number as in the case where all three components of  $\lambda_i$  are included.  

On the other hand, the bulk-boundary correspondence exists for surfaces that are perpendicular to the $(110)$ plane so that the electrons are necessarily confined to a plane which is not compatible with the spin-rotation axis used in the calculation of the mirror Chern number.  Nevertheless, we find that all calculations lead to correct approximate results as long as the axis of the spin-rotation symmetry is consistently chosen to be perpendicular to the plane where the electrons are confined to be.
In Fig. \ref{sfig10} we show that the surface Dirac cones are correctly reproduced in the low energy spectrum  of a system with $N=200$ layers stacked in the $(001)$ direction for  $\lambda_z=0.3$ eV and 
$\lambda_x=\lambda_y=0$. 

\section{Dependence of the step mode dispersion on the number of layers}
\begin{figure}[h!]
  \begin{center}  
    \includegraphics[width=  \textwidth]{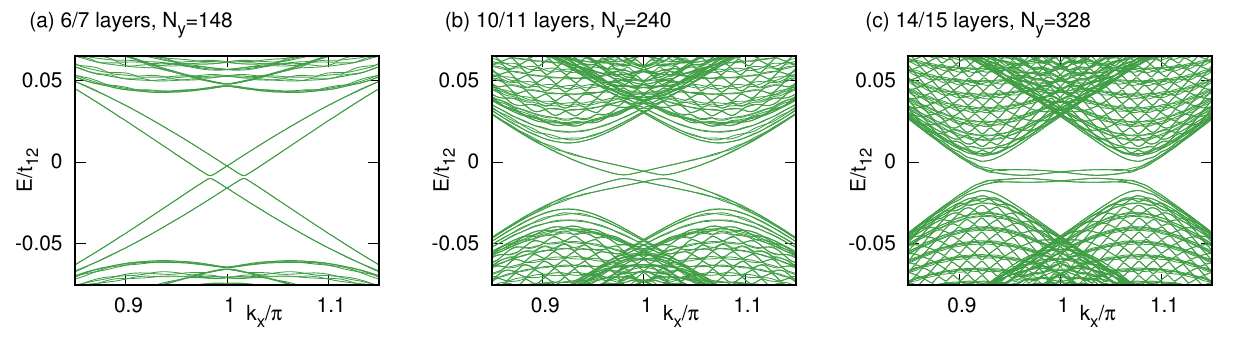}
  \vspace{-0.5	cm}
  \end{center} 
 \caption{Dispersion of the step modes in the case of a monoatomic step formed between (a) 6 and 7 layers, (b) 10 and 11 layers, and (c) 14 and 15 layers. $N_y$ is the width of the sample used in each plot. }
\label{sfig1} 
\end{figure}

As shown in Fig.~\ref{sfig1}, for increasing number of layers $N$ the system tends to develop Dirac cones with closing of the bulk gap that are consistent with the $(001)$ surface states of the TCI. The dispersion of the step modes connecting these cones becomes more and more flat with increasing $N$. 

\section{Completeness of the topological description of the step modes}

For large number of layers $N$ one expects that number of the step modes appearing at an atomic step should depend only on the height of the step rather than on $N$ itself. This is because adding or substracting a layer far away from the surface should not affect the step states. Having this in mind it is possible to argue that the step states cannot be entirely determined by the mirror- and spin-resolved Chern numbers of the multilayer systems. By looking the spin-resolved Chern numbers given in Fig. 2 of the main text we see that it is possible that two different even-layer systems have the same spin-resolved Chern numbers (e.g. $N=20$ and $N=22$) and the same mirror-resolved Chern numbers (always zero), and we expect that such situation  can occur for arbitrarily large $N$. We can now make an interface of these systems so that there is a unit height step on top and bottom surfaces. 
Such kind of system must  support step modes because using the arguments discussed in the main text we know that for large $N$ the unit height steps support step modes and the surfaces are only weakly coupled in this limit. 
Thus it remains an open question whether one can define additional topological invariant that would describe completely all possible interfaces. 

\section{Dependence of the energy gap on the Zeeman field}

\begin{figure}[h!]
  \begin{center}  
    \includegraphics[width=0.85  \textwidth]{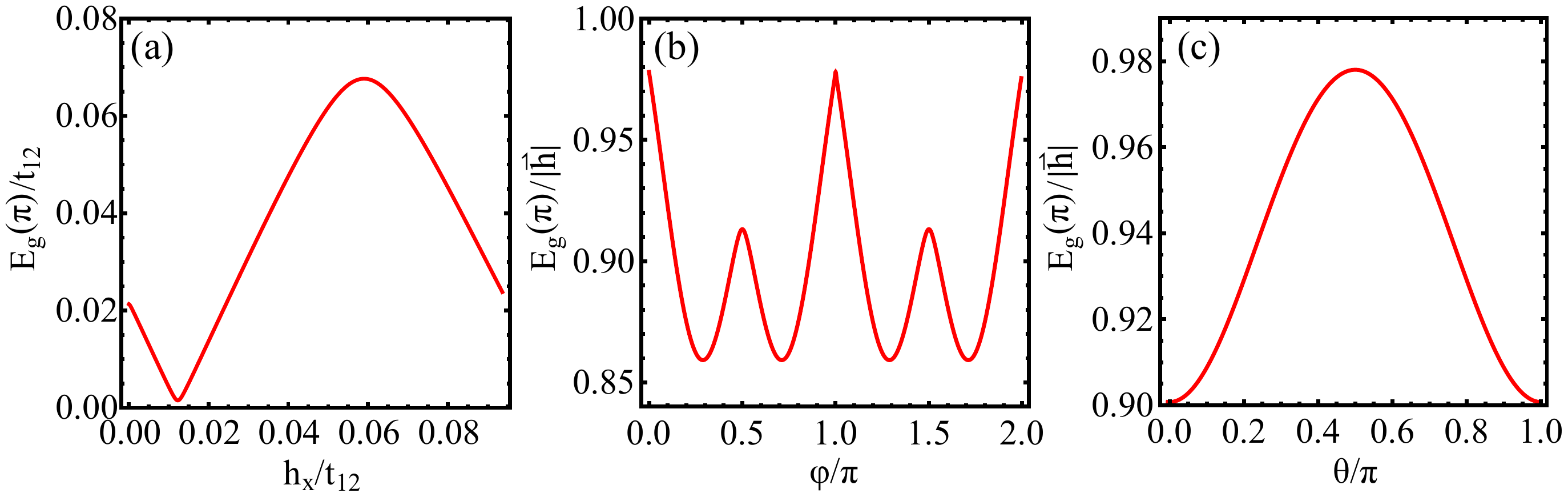}
  \vspace{-0.5	cm}
  \end{center} 
 \caption{ (a) Dependence of the energy gap $E_g$ at $k_x=\pi$ on the Zeeman field magnitude $h_x$. (b),(c) The dependence of $E_g$ on the direction of $\vec{h}$ within $xy$ plane given by the azimuthal angle $\varphi$ and $xz$ plane given by the polar angle $\theta$. The magnitude of the Zeeman field is $|\mathbf{h}|=0.025$ eV. In all figures the step is formed between regions with 2 and 3 layers, $\lambda_x=\lambda_y=\lambda_z=0.3$ eV and $N_y=52$.}
\label{sfig7} 
\end{figure}

The behavior of the energy gap  $E_g$ of a system with step states is non-trivial when the Zeeman field is switched on. In Fig.~\ref{sfig7}(a) we show the dependence of $E_g$ on $h_x$. Consistently with the results presented in the main text above some critical value of $h_x$ the gap opens up to maximal value around $h_x=0.06t_{12}$ but then it decreases and tends to close again. 
In our explanation the closing and reopening of the energy gap will not be seen experimentally because of the spontaneous magnetization present due to magnetic impurities/magnetic instability of step mode electrons.
The dependence on the direction of the field in the $xy$ and $xz$ planes is shown in Figs.~\ref{sfig7}(b-c). We expect that in the case of magnetic instability there exists an easy-axis along $x$-direction because the magnetization in $x$-direction maximizes the gap and leads to a minimum of the free energy.

\section{Dependence of the domain wall state energies  on the system parameters}
\begin{figure}[h!]
  \begin{center}  
    \includegraphics[width=0.65  \textwidth]{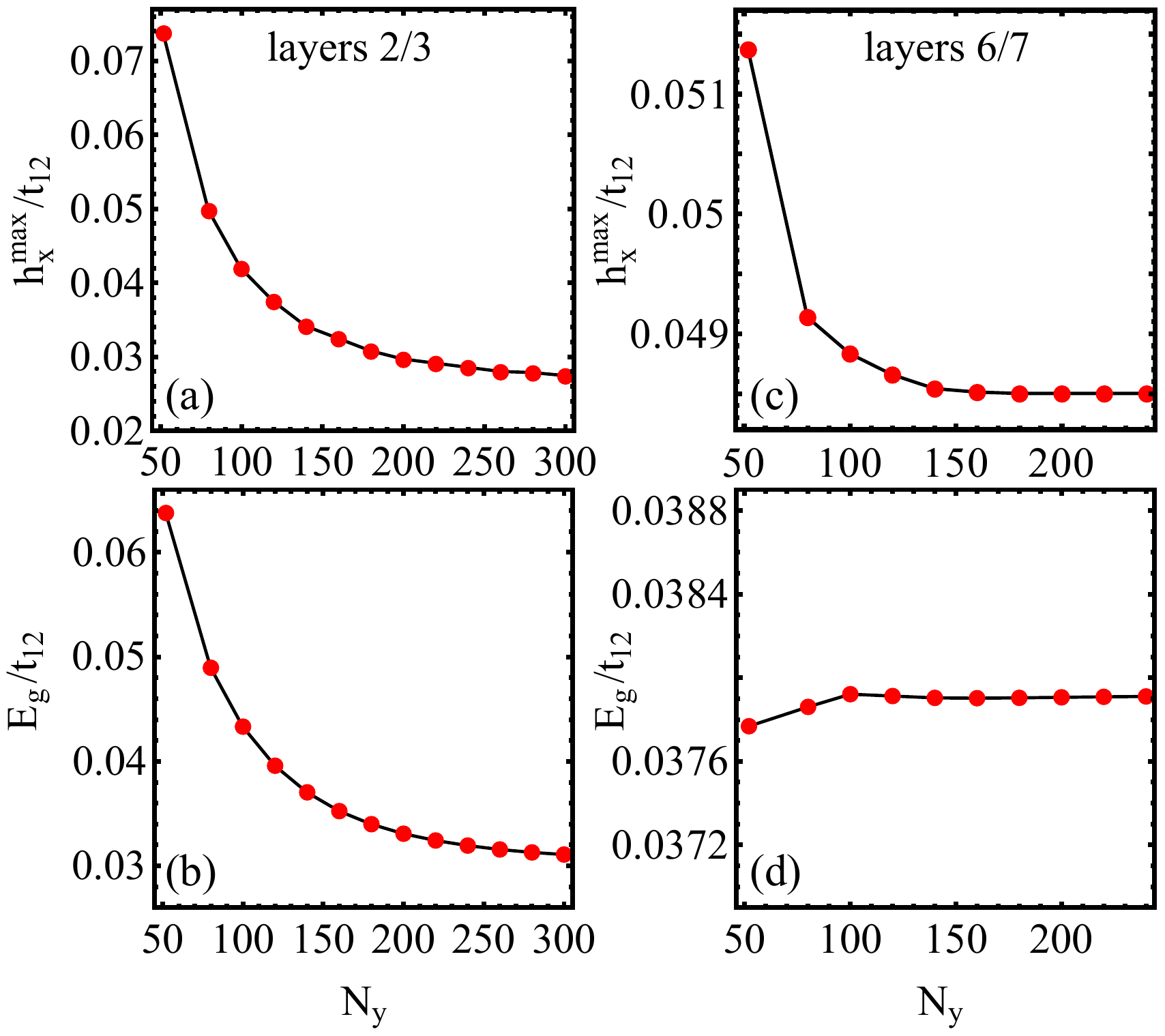}
  \vspace{-0.5	cm}
  \end{center} 
 \caption{Dependence of the Zeeman field $h_x^{\rm max}$ maximizing the energy gap and its maximal value $E_g$ on $N_y$. ($N_y/2$ is the spacing between the two steps.)  The steps are formed between regions of (a),(b) 2/3 layers  and (c),(d) 6/7 layers. We have used $\lambda_z=0.5$ eV and $\lambda_x=\lambda_y=0$.}
\label{sfig5} 
\end{figure}
\begin{figure}[h!]
  \begin{center}  
    \includegraphics[width=0.7  \textwidth]{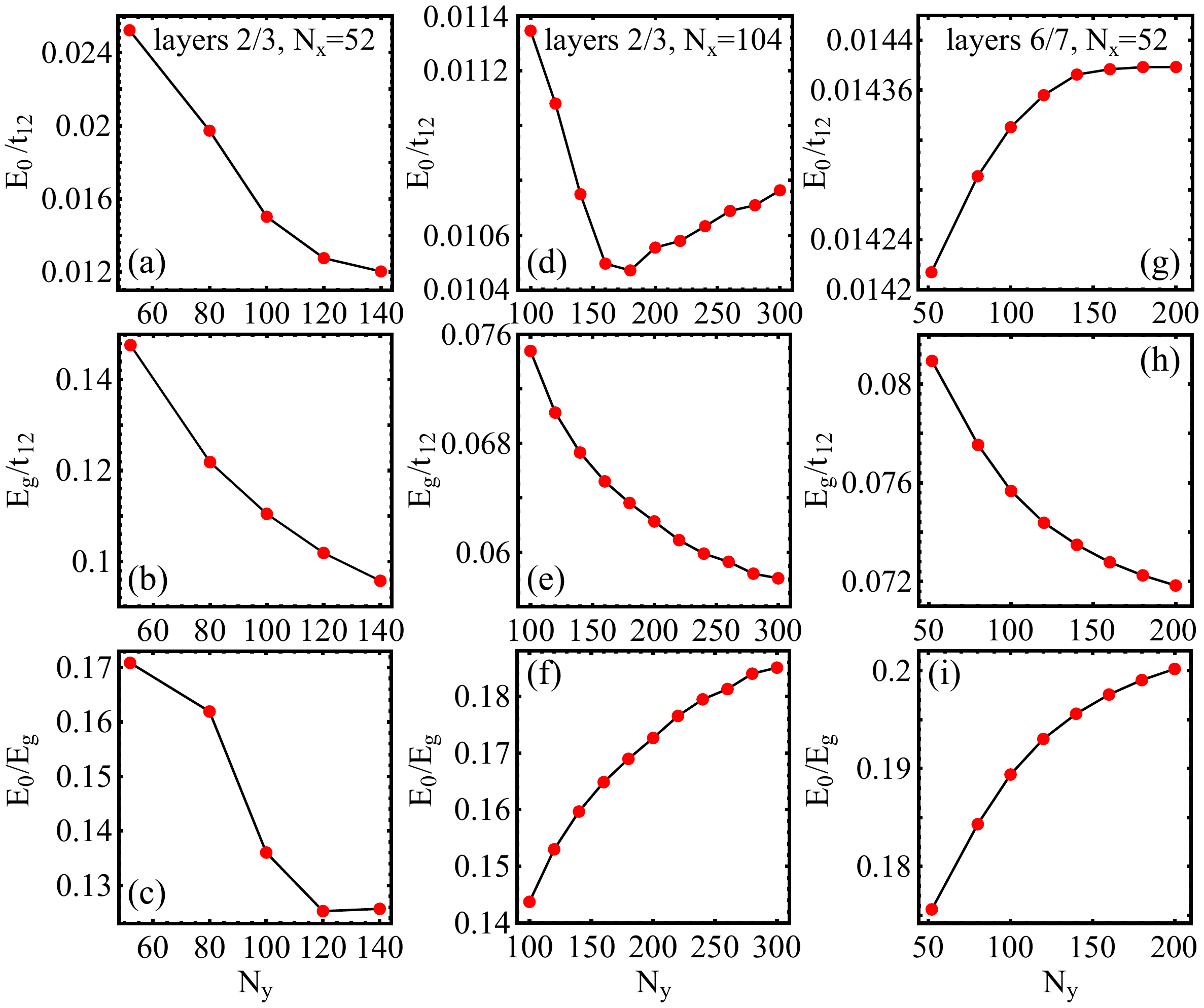}
  \vspace{-0.5	cm}
  \end{center} 
 \caption{The energy of bound states $E_0$, the energy gap $E_g$  and their ratio  $E_0/E_g$ as functions of  $N_y$. We have chosen $h_x=h_x^{\rm max}$ for each $N_y$ and considered a DW between regions with opposite magnetizations $h_x$ and $-h_x$. The energies are shown for system sizes (a),(b),(c) 2/3 layers and $N_x=52$, (d),(e),(f) 2/3 layers and $N_x=104$,  and (g),(h),(i)  6/7 layers and $N_x=52$. We have used $\lambda_z=0.5$ eV and $\lambda_x=\lambda_y=0$.}
\label{sfig6} 
\end{figure}
The scaling properties of the energy gap  of the step modes $E_g$ and the energy of the DW states $E_0$ in case of sharp DWs are complicated and depend on the value of the Zeeman field. Here we consider the field in the $x$ direction. In Fig. \ref{sfig5} we show the values of $h_x=h_x^{\rm max}$ that maximize  $E_g$ for systems of  2/3 and 6/7 layers as a function of $N_y$. We observe saturation of both  $h_x^{\rm max}$ and $E_g$ in the limit $N_y\to\infty$. We now consider the case of optimal energy gap and study the energies of the DW states for magnetic domains with opposite magnetization $\pm h_x$ as a function of the system size. The results are presented in  Fig. \ref{sfig6} for various system sizes and we consider $E_0/E_g$ as the figure of merit. In these figures $N_y$ is taken in a range of roughly  $N_x<N_y<5/2N_x$. We see that $E_0/E_g$ is typically between $0.1$ and $0.2$  and there is only a weak dependence on the system size.
\begin{figure}[h!]
  \begin{center}  
    \includegraphics[width=0.45  \textwidth]{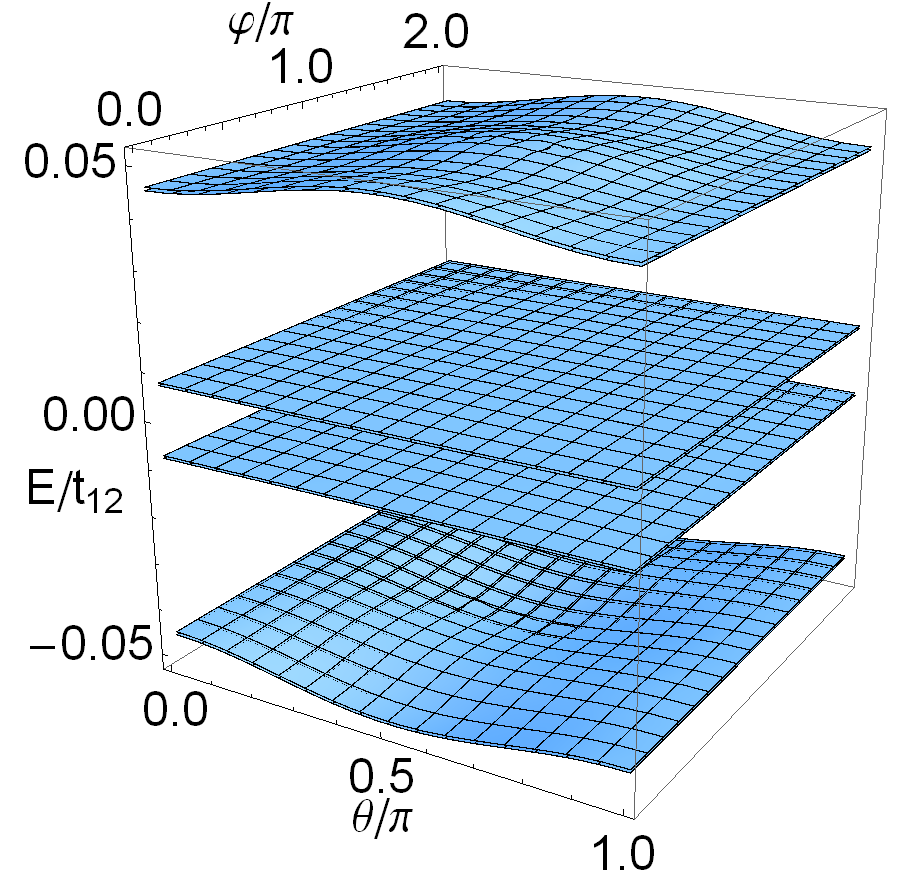}
  \vspace{-0.5	cm}
  \end{center} 
 \caption{Dependence of the low-energy spectrum for a system with two magnetic domains with magnetization $\pm \vec h$ on the magnetization direction given by azimuthal $\varphi$ and polar $\theta$ angles. The system consists of 2/3 layers with $N_x=N_y=52$, $|h|=t_{12}/30$ and  $\lambda_x=\lambda_z=0.4$ eV.}
\label{sfig8} 
\end{figure}
We also find that the energies $E_g$ and $E_0$ are surprisingly immune to the changes of the Zeeman field direction. In Fig.~\ref{sfig8} we show the low energy spectrum as function of the field direction for a fixed  magnitude of the Zeeman field.

\section{Low-energy theory for the step modes and domain walls}
By solving the eigenstates corresponding to one of the steps at $k_x=\pi$ and by expanding the Hamiltonian for momentum $k_x=\pi+k$, the low-energy theory for the step modes in the absence of Zeeman field can be written as
\begin{equation}
{\cal H}_{\rm step}(k)=\Delta \sigma_0 \tau_x + v k \sigma_z \tau_0+\tilde{v} k \sigma_0 \tau_z+\Lambda \sigma_y \tau_z+\tilde{\Lambda} k \sigma_x \tau_y,
\end{equation}
where $\Delta$ is the energy of the step modes at $k=0$ (in the absence of $\lambda_x$), $v$ and $\tilde{v}$ describe the velocities of the step modes ($|v|>|\tilde{v}|$),  and $\Lambda$ and $\tilde{\Lambda}$ are small terms arising due to the spin-orbit coupling $\lambda_x$. We find that in the lowest order approximation $\lambda_y$ does not have any effect in the low-energy theory. If $\lambda_x=0$  the model obeys spin-rotation symmetry around $z$-axis 
\begin{equation}
S_z {\cal H}_{\rm step}(k) S_z = {\cal H}_{\rm step}(k), \ S_z=\sigma_z \tau_0 \ \rm{for} \ \Lambda=\tilde{\Lambda}=0,
\end{equation}
and supports two pairs of gapless counterpropagating edge modes as described in the main text. Due to $\lambda_x \ne 0$ ($\Lambda \ne 0$, $\tilde{\Lambda}\ne0$) the spin-rotation symmetry is weakly broken and the step modes are gapped.

Within this low-energy model the non-symmorphic symmetries become usual symmorphic symmetries, which are obtained by projecting them to the states at $k_x=\pi$. The non-symmorphic chiral symmetry defines a chiral symmetry
\begin{equation}
C{\cal H}_{\rm step}(k)C=-{\cal H}_{\rm step}(k), \ C=\sigma_y \tau_y \label{chiral_low-energy}
\end{equation}
and the $k$-dependent mirror symmetry leads to a relation
\begin{equation}
M_x {\cal H}_{\rm step}(-k) M_x =  {\cal H}_{\rm step}(k), \ M_x=\sigma_x \tau_x. 
\end{equation}
Additionally the system obeys time-reversal symmetry 
\begin{equation}
\sigma_y \tau_x {\cal H}^*_{\rm step}(-k) \sigma_y \tau_x={\cal H}_{\rm step}(k).
\end{equation}
This is the most general linearized Hamiltonian in the presence of these symmetries.

\begin{figure}[h!]
  \begin{center}  
    \includegraphics[width=0.5  \textwidth]{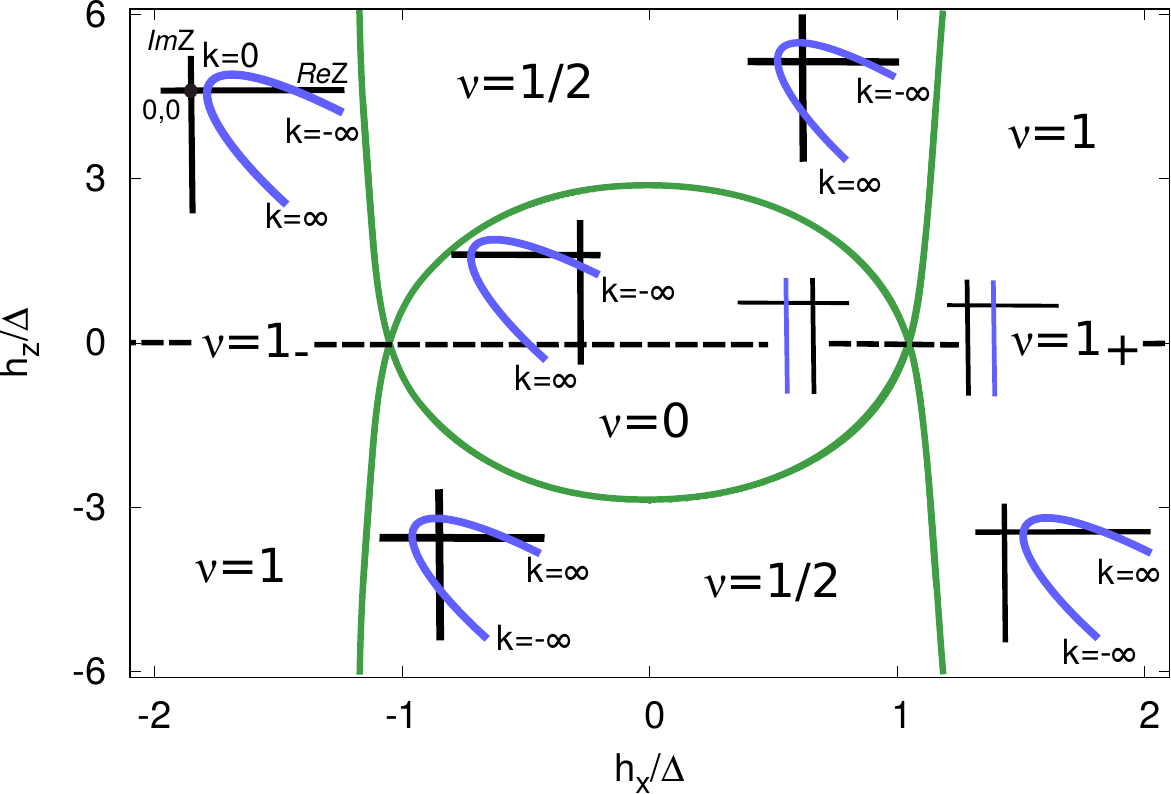}
  \vspace{-0.5	cm}
  \end{center} 
 \caption{Topological phase diagram of the low-energy model for the step modes  as function of the Zeeman field components $h_x$ and $h_z$ for
$\Delta=0.003$ eV, $v=-0.114$ eV, $\tilde{v}=-0.019$ eV, $\Lambda=0.000032$ eV, $\tilde{\Lambda}=0.01$ eV and $c=0.88$.}
\label{sfig_topological_phase_diagram} 
\end{figure}

The Zeeman field component $h_x$ preserves both $M_x$ and $C$, whereas $h_z$ breaks $M_x$ but preserves $C$. Therefore, we consider the effect of both of these components on the topology of the system. In the low-energy theory around $k_x=\pi$ they give rise to a Hamiltonian
\begin{equation}
{\cal H}^Z_{\rm step}= h_z \sigma_z \tau_0 + c h_x \sigma_x \tau_0,
\end{equation}
where the factor $c$ takes into account the renormalization of the effect of $h_x$ within the low-energy theory and $h_z$ is practically unmodified due to the fact that the step modes are spin-polarized as a good approximation. Since both of these terms obey the chiral symmetry [Eq.~(\ref{chiral_low-energy})], we can utilize the chiral symmetry to block-off diagonalize the Hamiltonian
\begin{equation}
U^\dag \big[ {\cal H}_{\rm step}(k) + {\cal H}^Z_{\rm step} \big] U= \begin{pmatrix} 0 & D(k) \\ D^\dag(k)& 0\end{pmatrix}, \ U=\begin{pmatrix} \tau_z & -\tau_z \\ \tau_x&  \tau_x\end{pmatrix}, \ D(k)=\begin{pmatrix} (-i \tilde{\Lambda}   - v - \tilde{v} )k -h_z & \Delta+c h_x - i \Lambda  \\ \Delta - c h_x - i \Lambda &  (-i \tilde{\Lambda}  - v  +\tilde{v}) k -h_z \end{pmatrix}.
\end{equation}
It is easy to see ${\rm Im}\{\det[D(k)]\}$ is independent of $h_x$ and $h_z$ at $k=0$ and $k \to \pm \infty$. Moreover, the sign of  ${\rm Im}\{\det[D(k)]\}$ is different at $k=0$ and $k \to \pm \infty$ [see Fig.~\ref{sfig_topological_phase_diagram}], and in the following we consider 
\begin{equation}
{\rm Im}\{\det[D(k \to 0)]\} >0 \ \rm{and} \  {\rm Im}\{\det[D(k \to \pm \infty)]\} < 0 \label{boundary-condition}
\end{equation}
\textit{as a constraint which is always required to be satisfied}. Within the low-energy theory this requirement is robustly satisfied even if we introduce an arbitrary small perturbation to the Hamiltonian. Moreover, our considerations are valid beyond the low-energy theory because if we equate the $k \to \pm \infty$ limit with $k \to 2 \pi$  this constraint is always satisfied  in the full theory where all the higher order terms in the expansion of ${\cal H}_{\rm step}(k)$ are correctly taken into account. This follows from the fact that the chiral symmetry $C$ originates from non-symmorphic chiral symmetry, which guarantees that a relation ${\rm Im}\{\det[D(0)]\}=-{\rm Im}\{\det[D(2\pi)]\}$ should always be satisfied \cite{Shiozaki2015supp}.  Importantly, in the presence of the constraint (\ref{boundary-condition})
 we can identify topologically distinct phases by plotting the trajectories of the $Z(k)=\det[D(k)]$ in the complex plane  [see Fig.~\ref{sfig_topological_phase_diagram}].

If we first set $h_z=0$ we reproduce the phase diagram discussed in the main text. Here the non-symmorphic chiral $\mathbb{Z}_2$ invariant $\nu$ can be determined by plotting the trajectory of $Z(k)$ from $k=0$ to $k=\infty$ in analogy to the trajectory plotted from $k=0$ to $k=2 \pi$ in the case of lattice Hamiltonian obeying non-symmorphic chiral symmetry \cite{Shiozaki2015supp}.  In the case of $\nu=0$ the trajectory is on the left side of the origin and in the case $\nu=1$ on the right side of the origin [see Fig.~\ref{sfig_topological_phase_diagram}]. Therefore, these two phases cannot be smoothly deformed into each other without going through the origin (i.e.~closing the energy gap). Moreover, the trajectory of $Z(k)$ from $k=-\infty$ to $k=0$ reproduces the same trajectory so that the behavior of the Hamiltonian at the momenta $k \to\pm \infty$ can identified with the  behavior of the lattice Hamiltonian at momentum $k=2\pi$, and therefore we expect that the full Hamiltonian where all the higher order terms in the expansion of ${\cal H}_{\rm step}(k)$ are correctly taken into account would also reproduce the same behavior. Therefore, although we have identified the $\nu=0$ and $\nu=1$ phases using the low-energy theory we expect that they are 
topologically distinct phases also in the case of the full lattice Hamiltonian (in analogy to the topologically distinct phases discussed in Ref.~\cite{Shiozaki2015supp}).  As discussed in the main text the band-inversions at $h_x=\pm h_c$ occur at different mirror sectors so that we can additionally label the non-trivial phases as $\nu=1_\pm$ as long as $h_z=0$.

The $z$-component of the Zeeman field $h_z$ breaks the mirror symmetry but preserves the chiral symmetry. 
Therefore for $h_z \ne 0$ the trajectories $k=0$ to $k=\infty$ and $k=-\infty$ to $k=0$ are no longer identical. However, as long as both of these trajectories remain on the same side of the origin, the Hamiltonians are topologically equivalent with the corresponding $\nu=0$ (trajectory on the left side) or $\nu=1$ (trajectory on the right side) Hamiltonians discussed above. Interestingly, in the presence of $h_z \ne 0$ there exists also a third topologically distinct phase within the low-energy theory, which we have labelled $\nu=1/2$. In this case the trajectories from $k=-\infty$ to $k=0$ and $k=0$ to $k=\infty$ are on the different sides of the origin (so that in a sense half of the trajectory is on the left and half on the right). This phase does not have lattice analogue \cite{Shiozaki2015supp} but it is still a topologically distrinct phase within the low-energy theory, because there necessarily exists a gap closing when deforming the Hamiltonian belonging to this topological class into a Hamiltonian belonging to either $\nu=0$ or $\nu=1$ topological classes in the presence of the constraint (\ref{boundary-condition}).  

This topological phase diagram enables a full description of the properties of the magnetic DWs where the magnetization $(h_x(x), 0, h_z(x))$ varies slowly as a function of $x$. For such kind of smooth magnetic DWs the number of gap closings (zero energy DW states) is determined by the number of topological phase transitions occurring along the trajectory $(h_x(x), 0, h_z(x))$ as a function of $x$. For example if one considers a DW between $\nu=1_+$ and $\nu=1_-$ phases any smooth DW will lead to at least two gap closings. The low-energy theory does not allow us to calculate the DW states appearing in the case of sharp DWs where the magnetization $(h_x(x), 0, h_z(x))$ varies quickly as a function of $x$. However, the numerical calculations described the main text and in the supplementary material are consistent with the idea that the number of low-energy bound states appearing in the case of sharp DWs is the same as the number of gap closing points in the case of smooth DWs.

\section{Calculation of the tunneling conductance}

\begin{figure}[b!]
  \begin{center}  
    \includegraphics[width=0.5  \textwidth]{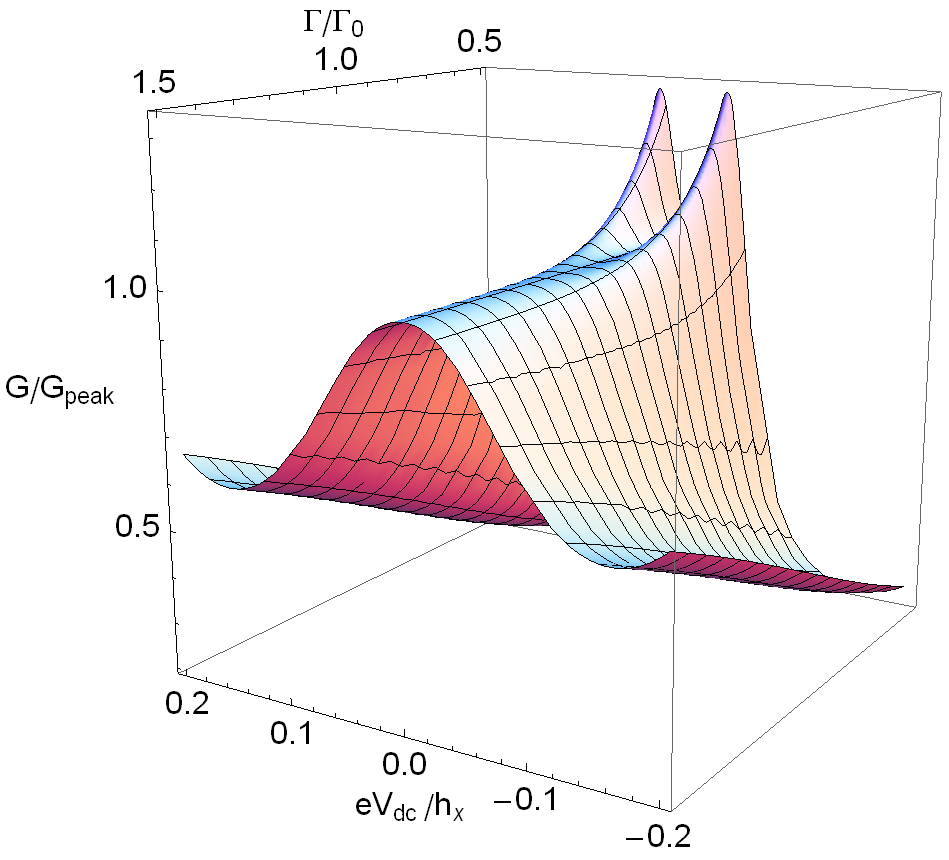}
  \vspace{-0.5	cm}
  \end{center} 
 \caption{Evolution from zero-bias peak to double peak structure as a function of $\Gamma$. We have used the energy spectrum shown in the main text in Fig.~3(c).}
\label{sfig99} 
\end{figure}

The tunneling current $I$ due to applied voltage $V_{dc}$ is described by \cite{Bruus}
\begin{equation}
 I(V_{dc})=e \int_{-\infty}^{\infty}dE \ T(E)[n_F(E)-n_F(E+eV_{dc})],
\end{equation}
where $n_F$ is Fermi function and $T(E)$ the transmission probability given by
\begin{equation}
 T(E)=\frac{2\pi}{\hbar}\sum_{\nu,\mu}|T_{\nu\mu}|^2A_{L\nu}(E)A_{R\mu}(E+eV_{dc}).
\end{equation}
Here $A_{L\nu}(E)$ and $A_{R\mu}(E)$ are the spectral functions of states $\nu$ and $\mu$ localized on the left $L$ and right $R$ side of the tunneling barrier, and $T_{\nu\mu}$ is the tunneling matrix element. 
In the case of soft point-contact spectroscopy we may assume that one of the systems has a constant density of states so that
\begin{equation}
\frac{2\pi e}{\hbar}\sum_{\nu}|T_{\nu\mu}|^2A_{L\nu}(E)\simeq const = I_0 
\end{equation}
and thus in zero-temperature limit 
\begin{equation}
 G=\frac{dI}{dV}=eI_0\sum_{\mu}A_{R\mu}(eV_{dc}).
\end{equation}
In realistic systems the spectral functions have a broadening $\Gamma$ due to electron-phonon and electron-electron interactions, disorder and coupling of the tip to the system. The simplest approximation in which these effects can be accounted for is Lorenzian form 
\begin{equation}
 A_{\mu}(E)=\frac{1}{\pi}\frac{\Gamma}{(E-\epsilon_\mu)^2+\Gamma^2},
\end{equation}
where $\epsilon_\mu$ is the energy of state $\mu$.
This way we obtain
\begin{equation}
 \frac{G(V_{dc})}{G_0}=\sum_{\mu}\frac{1}{(eV_{dc}-\epsilon_\mu)^2/\Gamma^2+1}
\end{equation}
with $G_0={eI_0}/{\pi\Gamma}$, and this formula has been used for plotting the tunneling conductances in the main text. Moreover, we define $G_{\rm peak}=G(0)$ and $\Gamma_0=\sqrt{3} E_0$, and in the main text we use $\Gamma=\Gamma_0$ in all figures. Because the DW states are not exactly at zero-energy one would  observe a double-peak structure in the tunneling conductance in the case of small broadening. However, for broadening $\Gamma \geq \Gamma_0$ one observes a single zero-bias peak (see Fig.~\ref{sfig99}).

\section{Domain wall states in the presence of randomly arranged magnetic impurities}

\begin{figure}[h!]
  \begin{center}  
    \includegraphics[width=0.75  \textwidth]{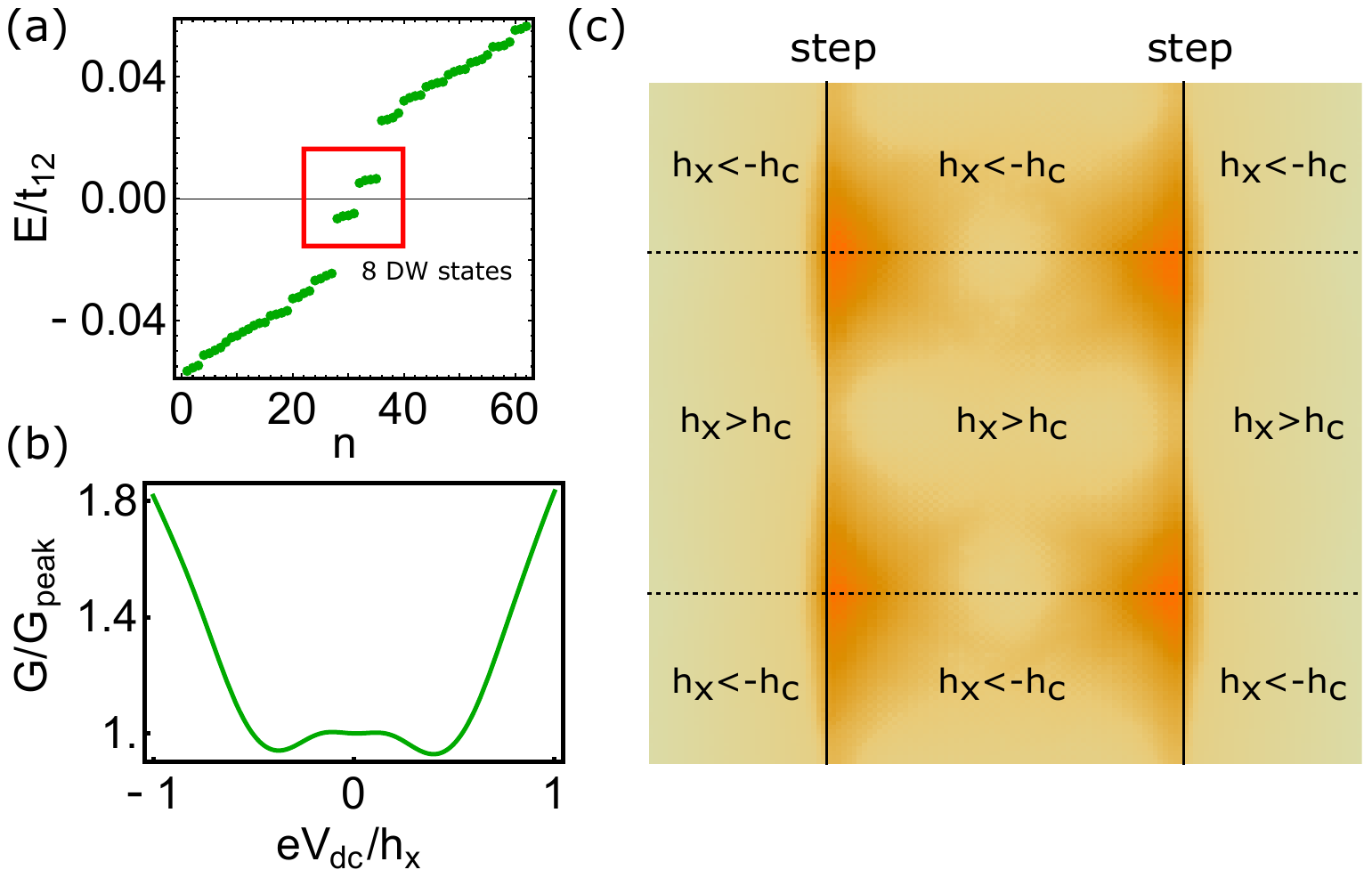}
  \vspace{-0.5	cm}
  \end{center} 
 \caption{(a), (b) Low-energy spectrum and differential conductance for a system containing two steps and  $10\%$ concentration of randomly arranged magnetic impurities in a domain wall configuration. The domain walls are created by choosing opposite $h_x$ in the two halves of the system.  (c) Corresponding LDOS for the DW states. The parameters are $N_x=160$, $N_y=140$,  $\lambda_z=0.5$ eV, $\lambda_x=\lambda_y=0$ and $|h_x|=0.34$ eV. The broadening used in the calculation of the differential conductance is $\Gamma=0.3 |h_x|$.} 
\label{sfig11}
\end{figure}

In the case when the magnetism at the step is due to the magnetic impurities one can ask whether the  disorder in the arrangement of these impurities will affect the low-energy states at the domain walls. To show that these states are not affected by the disorder we implement a system with  $10\%$ concentration of magnetic impurities distributed randomly in space with magnetic moments pointing along $x$. Then we create a domain wall by choosing opposite $h_x$ in two halves of the system with magnitude such that $0.1 |h_x|>h_c$ to compensate the dilution of magnetic centers. The results for the energy spectrum, differential conductance and LDOS [Figs. \ref{sfig11}(a-c)] are very similar to the ones obtained without disorder (Fig. 3 in the main text). 

\begin{figure}[h!]
  \begin{center}  
    \includegraphics[width=0.65  \textwidth]{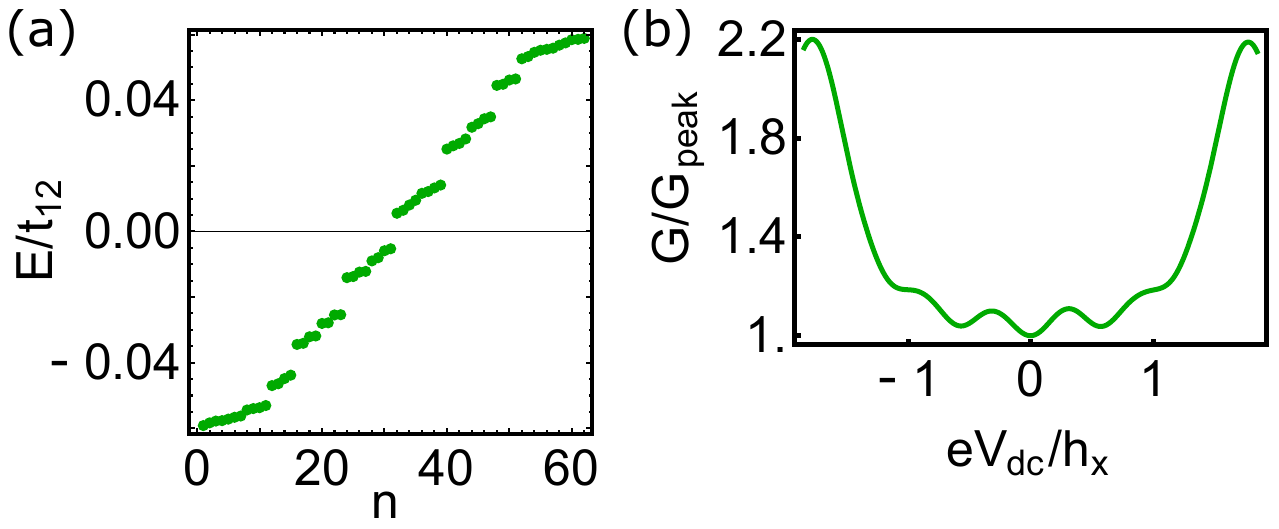}
  \vspace{-0.5	cm}
  \end{center} 
 \caption{(a), (b)  Low-energy spectrum and differential conductance for a system containing two steps and  $10\%$ concentration of randomly arranged magnetic impurities where the magnetic moments are oriented randomly in the (x,y)-plane.
The parameters are $N_x=160$, $N_y=140$,  $\lambda_z=0.5$ eV, $\lambda_x=\lambda_y=0$ and $|\vec{h}|=0.34$  eV. The broadening used in the calculation of the differential conductance is $\Gamma=0.3 |\vec{h}|$.} 
\label{sfig12}
\end{figure}

Recent point-contact experiments done on the SnTe class of materials  show not only zero-bias conductance peaks at low temperature, but also a soft gap feature at higher temperatures \cite{Das2016-supp,Mazur2017-supp}.  Within our theoretical framework we expect that the main effect of temperature will be melting of the magnetic order. To simulate this we assume that the randomly distributed magnetic impurities also have random direction of the magnetic moments in the (x,y)-plane. This leads to the energy spectrum and differential conductance shown in Fig. \ref{sfig12}. As we can see the soft gap feature is correctly reproduced.

\section{Competition between different types of correlated states at the steps and possible experimental signatures}

We emphasize that within the framework of our theoretical investigations the correlated phase realized in the experiments \cite{Das2016-supp,Mazur2017-supp} does not necessarily need to be an easy-axis ferromagnetic. Rather we use the easy-axis ferromagnet as an example of non-superconducting state which is consistent with the parametric dependencies of the observed tunneling conductance -- demonstrating that the experimental observations can be explained without requiring the existence of superconductivity. In this section we briefly discuss the possible options for the correlated states realized at the steps and the experimental methods for distinguishing them from each other.

The appearance of correlated states at the steps due to the approximately flat bands have analogies to several other systems supporting flat bands, such as symmetry-broken states (ferromagnets, exciton condensates and more complicated order) in quantum Hall systems \cite{Girvin99-supp, Moon95-supp, Nomura2006-supp, Alicea06-supp, Yang06-supp, Kharitonov2012-supp, Kharitonov2012-supp2, Young2013-supp}, competition between flat-band surface superconductivity and magnetism in semimetals \cite{Kopnin2011-supp, Volovik2018-supp, Ojajarvi18-supp, Pamuk-supp, Lothman-supp} and  the correlated states in graphene Moir\'e superlattices \cite{Cao-supp, Cao-supp2, Sharper2019-supp, Efetov2019-supp, Liu2019-supp}. The generic feature of these systems is that they are often characterized by a competition between different types of order parameters which are almost degenerate with each other \cite{Nomura2006-supp, Alicea06-supp, Yang06-supp} and the ground state depends on the details of microscopic model \cite{Kharitonov2012-supp, Ojajarvi18-supp}. Therefore, it is practically impossible to predict what type of correlated state is realized in the flat-band systems by using purely theoretical methods. Nevertheless, the combination of theoretical and experimental investigations has lead to impressive progress in our understanding of these systems \cite{Kharitonov2012-supp, Kharitonov2012-supp2, Young2013-supp}. There the idea is essentially to systematically analyze the properties of the possible candidate states and to propose experimental tests to find out which of them is realized. In the case of step states, by relying on the analogy to the previously studied systems we expect that both easy-axis ferromagnet (flat-band ferromagnetic state caused by the Coulomb interactions) and flat-band superconductivity (e.g. due to phonon-mediated attractive interactions) are realistic possibilities \cite{Girvin99-supp, Moon95-supp,Ojajarvi18-supp}. However, since there exists four approximately flat bands also more complicated order might appear in analogy to the SU(4)-symmetry-broken states studied in graphene in the quantum Hall regime \cite{Nomura2006-supp, Alicea06-supp, Yang06-supp, Kharitonov2012-supp, Kharitonov2012-supp2, Young2013-supp}.

Therefore we need to find out which of the candidate states are consistent with the experimental observations. As we discussed above the intrinsic properties of the non-superconducting symmetry-broken states at the step defects can lead to parametric dependencies of the  tunneling conductance  which can mimic all the features which are often attributed to the unconventional superconductivity and Majorana modes \cite{Das2016-supp, Mazur2017-supp, Wang2016-supp, Aggarwal2016-supp}. This means that
the current experimental evidence is not conclusive and new experiments are necessary. Four-probe STM and spin-polarized STM measurements can determine whether the correlated state at the steps is related to superconductivity or magnetic order. Moreover, our theory predicts that the topological nature of the step modes is independent of the thickness of the sample. Therefore, the correlated states can be studied also in thin films, where the dispersion of the step modes can be controlled with the thickness and the density with the help of gate voltages. This makes it possible to study the competition between different phases \cite{Ojajarvi18-supp, Lothman-supp} in a controllable way.

There are several indications that our approach might capture the essential physics related to the experiments \cite{Das2016-supp, Mazur2017-supp, Wang2016-supp, Aggarwal2016-supp}. In particular, it provides a possible explanation why the zero-bias anomalies appear in topologically nontrivial multilayer systems (topological step modes appear only if the systems on the different sides of the step are topologically distinct) and a microscopic mechanism for the appearance of symmetry-broken states in materials where no spontaneous symmetry-breaking happens in the bulk (steps support approximately flat bands with a large density of states). Moreover, our theory explains why these zero-bias anomalies are observed in materials with a sublattice degree of freedom (this is the origin of the non-symmorphic chiral symmetry). Since the explanation does not require the existence of superconductivity it is also consistent with the lack of direct signatures of superconductivity \cite{Mazur2017-supp} and it provides a possible explanation for the observation that the increase of the concentration of magnetic dopants enhances the effect  \cite{Mazur2017-supp} (magnetic instability instead of superconducting one). From the viewpoint of magnetic instability the magnetic impurities give rise to another mechanism (in addition to the Coulomb interactions between electrons) favouring magnetic order and therefore if there exists competition between different phases they help to stabilize the magnetic order. On the other hand, magnetic impurities are practically always destructive for the superconducting order both in conventional and unconventional superconductors \cite{Balatsky06-supp, Alloul09-supp, Mackenzie98-supp}, and also from the random matrix theory viewpoint the disorder in this symmetry class of superconductors would lead to subgap states at arbitrarily low energies and therefore to the destruction of the superconducting gap.


\begin{thebibliography}{39}
\bibitem [{foo({\natexlab{a}})}]{footnote0}%
  \BibitemOpen
  \bibinfo {note} {For small systems the tight-binding parameters can deviate
  significantly from the bulk values and lead to different topological
  properties \cite{Schmidt12018}. However, we consider small systems only for
  illustration purposes. Our results are based on symmetry and topological
  considerations which are valid for sufficiently large systems ($\gtrsim10$
  layers) where the bulk values of the parameters can be used.}\BibitemShut
  {Stop}%
\bibitem [{foo({\natexlab{c}})}]{footnote1}%
  \BibitemOpen
  \bibinfo {note} {Notice that this reasoning is valid also if the states are
  localized in the vicinity of the surface in the $z$-direction even if the
  system itself is infinite.}\BibitemShut {Stop}%
\bibitem [{foo({\natexlab{d}})}]{footnote2}%
  \BibitemOpen
  \bibinfo {note} {Due to bulk-boundary correspondence there will be $|\Delta
  C_\uparrow|$ step modes with spin $\uparrow$ propagating in one direction and
  $|\Delta C_\downarrow|=|\Delta C_\uparrow|$ step modes with spin $\downarrow$
  propagating in the opposite direction.}\BibitemShut {Stop}%
\bibitem [{foo({\natexlab{e}})}]{footnote4}%
  \BibitemOpen
  \bibinfo {note} {Also other types of order parameters can naturally support
  the appearance of domain walls, and the magnetic order is considered here
  only as an example.}\BibitemShut {Stop}%
\end{thebibliography}

%

 \end{document}